\documentclass[useAMS,usenatbib]{mn2e}
\usepackage{graphicx,epsfig,psfig,multicol}
\usepackage[normalem]{ulem}
\usepackage[dvipsnames]{xcolor}
\usepackage[]{inputenc,amssymb}
\usepackage{amsmath}
\usepackage{color}
\usepackage{epsfig}
\usepackage{textcomp}
\usepackage{xcolor,cancel}

\definecolor{webgreen}{rgb}{0,.5,0}
\definecolor{webbrown}{rgb}{.6,0,0}
\usepackage[pdfpagelabels]{hyperref}
\hypersetup{%
   colorlinks=true,hyperfootnotes=false,%
   breaklinks=true,%
   plainpages=false, bookmarksnumbered, bookmarksopen=true,
   bookmarksopenlevel=1,%
   urlcolor=webbrown, linkcolor=RoyalBlue, citecolor=webgreen,
   }

\usepackage[toc,page]{appendix}

\def \be{\begin{equation}}
\def \ee{\end{equation}}
\def \bea{\begin{eqnarray}}
\def \eea{\end{eqnarray}}
\def\etal{{et al.}}

\title[How outflows propagate through hot halo]{Long way to go: how outflows from large galaxies propagate through the hot halo gas}
\author[Sarkar, Nath, Sharma, Shchekinov]
{Kartick Chandra Sarkar$^{1,2}$ \thanks{kcsarkar@rri.res.in}, Biman B. Nath$^1$, Prateek Sharma $^2$, Yuri Shchekinov$^3$ \\\\
$^1$ Raman Research Institute, Bangalore, India, 560080 \\
$^2$ Department of Physics \& Joint Astronomy Program, Indian Institute of Science, Bangalore, India, 560012\\
$^3$ Southern Federal University, Rostov-on-Don, Russia
}

\begin{document}


\maketitle

\label{firstpage}

\begin{abstract}
Using hydrodynamic simulations, we study the mass loss due to supernova-driven outflows from Milky Way type disk galaxies,
 paying particular attention to the effect of the extended hot halo gas.
We find that the total mass loss at inner radii scales roughly linearly with total mass of stars formed,
and that the mass loading factor at the virial radius can be several times its value at inner radii  because of the swept up hot halo gas.
The temperature distribution of the outflowing material in the inner region ($\sim $10 kpc) is
 bimodal in nature, peaking at $10^5$ K and $10^{6.5}$ K, responsible for optical and
X-ray emission, respectively. The contribution of cold/warm gas with temperature $\le 10^{5.5}$ K
to the outflow rate within 10 kpc is $\approx 0.3\hbox{--}0.5$. The warm mass loading factor, $\eta_{3e5}$
($T\le 3 \times 10^5$ K) is related
to the mass loading factor at the virial radius ($\eta_{v}$) as $\eta_{v} \approx 25\, \eta_{3e5}\, \left(\mbox{SFR} /
{\rm M}_\odot{\rm yr}^{-1} \right)^{-0.15}$ for a baryon fraction of 0.1 and a starburst period of 50 Myr.
 We also discuss the effect of multiple bursts that are separated by both short and long periods.
The outflow speed at the virial radius is close to the sound speed in the hot halo, $\lesssim 200$ km s$^{-1}$. We
identify two `sequences' of outflowing cold gas at small scales: a fast ($\approx 500$ km~s$^{-1}$) sequence, driven by the unshocked free-wind; 
and a slow sequence ($\approx \pm 100$ km~s$^{-1}$) at the conical interface of the superwind and the hot halo.
\end{abstract}


\section{Introduction}
\label{sec:introduction}
Galaxies do not evolve as closed systems, and the amount and nature of infall and outflow regulate the crucial 
aspects of galactic evolution. The movement of gas in and out of a galaxy plays a crucial role in dictating the
star formation history of  the galaxy, which in turn determines other aspects of its evolution. The infall and
outflow of gas also shape the so-called galaxy eco-system, in the immediate vicinity of the galaxy. Not only
the evolution of the galaxy itself, but galactic outflows also have cosmological importance because they
enrich the intergalactic medium (IGM) with metals. The infall of the IGM gas into a galaxy depends, among 
other things, on the cooling efficiency of this gas, which in turn depends on the efficiency of outflows in 
depositing mass and metals. Galactic outflows, and in particular the amount of mass lost, are therefore important
links between galactic and cosmological evolution. This requires a detailed study of their dynamics up to
the virial radius.

The mass loss rate in outflows  has been estimated in various ways in the literature. In the standard scenario, 
outflows are believed to be excited mostly through the effect of multiple supernovae (SNe) arising from 
vigorous star formation in a galaxy. Recently other possible mechanisms, such as radiation pressure 
on dust grains embedded in the outflowing gas, and cosmic rays have also been invoked in launching these outflows. 
In the SNe driven scenario, \cite{larson74} estimated the total mass lost by equating the total thermal energy 
deposited in the interstellar medium (ISM) by multiple SNe, to the escape energy of the 
outflowing gas in a galaxy (which depends on the total mass, baryonic mass and the size of the galaxy). 
Equivalently, the mass lost equals the total thermal energy of the ISM divided by the square of the wind speed, 
which is likely of order the sound speed of the hot gas. This led to an estimate that a galaxy of mass (all 
baryonic) $\sim 5 \times 10^9$ M$_{\odot}$ would lose half of its mass in an outflow, and larger galaxies would 
lose relatively less mass. This idea led \cite{dekel86} to consider the effect of such winds in the evolution 
of dwarf galaxies, and they found that outflows from halos with virial speed less than $\sim 100$ km s$^{-1}$ 
have sufficient energy to eject most of the halo gas. This result suggested a dividing line between bright and 
diffuse dwarf galaxies (see also \cite{babul92}). If the outflow speed is comparable to the escape speed (which 
scales  with the disk  rotation speed, $v_c$), then it also means that the ratio of mass loss rate to SFR
$\eta ( \equiv \dot{M}/{\rm SFR}) \propto v_c^{-2}$  for energy driven outflows.
Such estimates of mass (and metals) lost were used in the early, semi-analytical calculations for the 
enrichment of the IGM \citep{tegmark93, nath97, ferrara00a}. If other mechanisms such as radiation pressure 
should dominate, then it has been shown that the outflow would be momentum driven instead of being driven by 
energy, and that $\eta \propto v_c^{-1}$ \citep{murray05}.

However, as observations of typical multiphase outflows (such as in M82) suggest, the estimate of mass lost is 
likely to be more complicated than as outlined above. The outflowing material consists of gas at different 
temperatures (ranging from the X-ray emitting hot gas to a cold phase containing molecular gas), and the speed 
throughout the outflow is hardly uniform. The  multiphase temperature/density structure and dynamics is further 
complicated by the non-spherical morphology of the outflow, which is moulded into a biconical shape by the 
interaction with the stratified disk  material of the star forming galaxy. 

The multidimensional and multiphase nature of galactic outflows calls for a more detailed numerical modelling, especially 
because only certain phases at smaller scales are accessible to observations. For the IGM, 
however, the scales close to the virial radius are the most important. Therefore, it is essential to understand the relation between
outflows at various scales via controlled numerical simulations.

Numerical simulations have helped one to overcome the limitations of 1-D semi-analytical calculations. 
\cite{maclow99} simulated SNe driven superbubbles of various mechanical luminosities in disks embedded in 
dark matter halos, and determined their efficiency in driving outflows from low-mass galaxies. The central 
source of energy (and mass) injection had a constant luminosity maintained for 50 million years. They found 
the range of luminosities (related to the star formation rate) that can drive out gas, completely or partially, from galaxies
of different masses. They found that only a small fraction of the total gas mass was 
expelled, except in the smallest galaxy considered (with $10^6$ M$_{\odot}$ gas). However, because of the absence of the
hot halo gas in their simulations, the results  cannot be directly applied to higher mass galaxies ($M \gtrsim 10^{12}\, M_\odot$).
The hot gas density in the halo is expected to be non-negligible, $n \sim 10^{-4}$ cm$^{-3}$ (e.g., see \citealt{sharma12}),
for a Milky-Way mass galaxy, and therefore 
the halo gas must be included in  order to study outflows at  scales $> 10$ kpc.

Recent simulations with GADGET (which includes the 
halo gas) by \cite{hopkins2012} have compared the efficiency of different physical processes in driving 
outflows, and also determined some scaling relations for the mass lost as a function of the galactic mass and 
disk  column density. However, these simulations study the outflow within a region of 50 kpc, and cannot be treated 
as an estimate of mass outflow at the virial radius. Moreover, in these
simulations (unlike that of \cite{maclow99}), the outflows are linked to 
the SFR, which is in turn linked to the disk  gas density, and which ultimately depends on the galaxy mass.
It is crucial to understand the dependence of  the large scale outflows on the halo mass and the SFR, 
since cosmological simulations of IGM enrichment that are not capable of resolving galaxies, rely on 
such prescriptions.  E.g., \cite{oppenheimer06} used 
various scaling for the mass loading factor ($\eta$, the ratio of mass loss rate to the SFR) with the galactic 
mass or velocity dispersion of the halo ($\eta \propto \sigma^{-1}, \sigma^{-2}$) in order to match the 
simulation results with observations.

The mass loading factor estimated at the virial radius 
has important implications for understanding dynamical processes accompanying enrichment of the Universe. Recent observations 
of the extended circum-galactic haloes bearing metals up to distances 100 -- 200 kpc from the parent galaxy \citep{tumlinson2011,bordoloi2014,mathes2014} 
call for a long-term numerical study covering passage of the outflows through the virial threshold.

\cite{hopkins2012} also found that the outflowing gas is mostly dominated by warm gas ($10^4 < T< 4 \times 10^5$ 
K), which is consistent with the results of \cite{melioli13} who have simulated the outflow from M82. However, 
it would be important to know the mass distribution of gas at different temperatures as a function of SFR and 
galactic mass, and at different galacto-centric radii. This is needed to compare the theoretical/simulation 
results with observations which are limited to small distances. For example, \cite{martin99} found that the 
rate of mass loss in the form of gas at $\sim 1$ keV is comparable to the SFR for galaxies with low rotation 
speed ($\le 130 $ km s$^{-1}$). However, these estimates are relevant for small distances (typically 
$\sim 0.1$ of the virial radius) where the emission measure of the gas is large, and may not be useful if one 
considers the mass flux at the virial radius, which is required for studies of the enrichment of the IGM. 

Observationally, the estimates of the mass loading factor $\eta$ has focussed on $L_\ast$ galaxies at $z\le 1$, 
with stellar masses of a few times $10^{10}$ M$_{\odot}$. These estimates range between $\eta \sim 0.3\hbox{--} 
30$ in this galactic mass range for $z\sim 0.1$ \citep{rupke2013, heckman00, bouche2012, bolatto2013}. 
At high redshifts, the estimates appear to be similar to values at present epoch \citep{martin2012, newman2012}. 
The corresponding observed outflow speed ranges between $100\hbox{--}800$ km s$^{-1}$. In a semi-analytic 
calculation, \cite{lagos2013} has predicted that $\eta$ saturates below a circular velocity of $\sim 80$ km 
s$^{-1}$. However, their calculation is limited to 1-D dynamics and also to length scales of order the disk 
height.

In this paper we investigate outflow properties varying the star formation rate over a wide range, keeping
the ISM and halo properties fixed to the Milky Way values. Our study is similar in spirit
to that of \cite{maclow99}, but we include the effect of hot halo gas in constraining the outflowing gas.
We do not consider the effects of mass loading on the central conical outflow
from thermally evaporated clumps and pre-existing clouds in the ISM (before the launch of the outflow), as in 
\cite{suchkov96}. We have also investigated the relation 
between outflows in the cold phase and the total outflow rate at smaller scales and at the virial radius. Moreover, we investigate
the detailed kinematics of the cold/warm gas and relate it with observations. 

The paper is organised as follows. In $\S$2, we discuss the mass model of the simulated galaxy, setting an
equilibrium initial condition and selection of the injection parameters. In $\S$3, we describe the PLUTO simulation code and various settings that we use.
 In $\S$4, we present the results of our runs, in $\S$5,  we discuss some of the implications of our work and finally in $\S$6
 we summarise our key findings.

\section{Mass model of the Galaxy}
\subsection{Gravitational potentials}
To model the density distribution of the gas , we consider two gravitational potentials. For the disk, we use the 
Miyamoto $\&$ Nagai potential \citep{miyamoto75} (in cylindrical coordinates ($R,z$)), 
\begin{equation}
 \Phi_{\rm disk }(R,z) = - \frac{GM_{\rm disk }}{\sqrt{R^2+(a+\sqrt{z^2+b^2})^2)}}, \:\: (\, a,b \geq 0 \,)  \,
\end{equation}
where $a$ and $b$ represent the scale length and the scale height of the disk (mass $M_{\rm disk }$) respectively. 
For the dark matter, we use a modified form of the NFW dark matter (DM), which unlike the original
 NFW profile \citep{nfw96}, has  a core with a finite dark matter density at the centre.
 The potential is given as
\begin{equation}
 \Phi_{\rm DM} = - \left( \frac{GM_{\rm vir}}{f(c)\,r_s} \right) \frac{\log(1+\sqrt{R^2+z^2+d^2}/r_s)}{\sqrt{R^2+z^2+d^2}/r_s} \:\: (\, d \geq 0 ),
\end{equation}
where $ f(c) = \log(1+c)-c/(1+c) $ with $ c = r_{\rm vir}/r_s $ , the 
concentration parameter and $d$ is the core radius of the DM distribution (see appendix \ref{DM} for 
the DM density profile). $M_{\rm vir}$, $r_{\rm vir}$ and $r_s$ are the total mass of the galaxy (including DM), the virial radius
and scale radius respectively. A full list of parameters for the model galaxy is given in the Table \ref{massparameters}.

 \begin{table}

  \centering
  \begin{tabular}{ l c l} 
   \hline\hline 
  parameters & values\\[0.5ex]
   \hline
   $M_{\rm vir} ({\rm M}_{\odot}) $                    
   & $10^{12}$\\ 
   $M_{\rm disk } ({\rm M}_{\odot})$ 
   & $5 \times 10^{10}$ \\
   $T_{\rm vir}$ K 
   & $3 \times 10^6$ \\
   $r_{\rm vir}$ (kpc)              
   & $258$ \\
   $c$ 
   & $12$  \\ 
   $r_s $ (kpc)  
   & $21.5$ \\
   $a$ (kpc)
   & $4.0$ \\
   $b$ (kpc)
   & $0.4$ \\ 
   $d$ (kpc)
   & $6.0$ \\
   $c_{s,\sigma}$ (km s$^{-1}$) 
   & $20.8$ \\
   $\mathcal{Z}_{\rm disk }$ ($\mathcal{Z}_{\odot}$) 
   & $1.0$ \\
   $\mathcal{Z}_{\rm halo}$ ($\mathcal{Z}_{\odot}$)
   & $0.1$ \\
   $\rho_{c}(0,0)$ (m$_p$cm$^{-3}$)
   & $3.0$\\
   \hline
   $H_R$ (kpc)       
   & $ \sim 2.2$ \\
   $H_z$ (kpc) 
   & $ \simeq 0.2$ \\
   $M_{\rm WIM} ({\rm M}_{\odot})$
   & $7 \times 10^8$ &\\
   $\rho_{\rm hot}(0,0)$ (m$_p$cm$^{-3}$)
   & $1.1 \times 10^{-3}$ \\
    \hline
  \end{tabular}
     \caption{Parameters used in our simulations.
           }
 \label{massparameters}
 \end{table}

\begin{figure}
 \includegraphics[width=6.0cm, angle=-90 ]{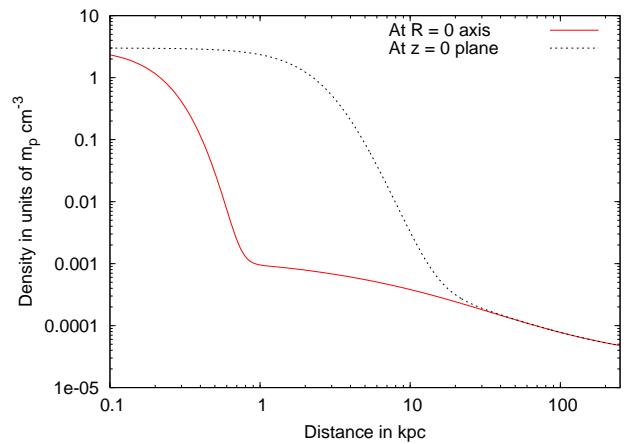}
 \caption {Gas density profiles for  the model galaxy along $R$ and $z$. 
           It shows that the disk component dominates at smaller $R$ and $z$, and at larger distances, the halo component dominates.
            }
 \label{gas_density}
\end{figure}
\subsection{Setting the initial density distribution}
\label{HE}
In our model, we consider two components of the interstellar medium (ISM): warm ionised medium (WIM), and a hot ionised
medium (HIM). Since the WIM (T $\sim 10^4$K, hereafter we call it the disk  gas) is a part of the disk  gas which is rotating
along with the stars, we consider the 
effect of rotation on the density distribution of the WIM. The HIM, however, according to the galaxy formation theory,
traces out the potential of the dark matter with no net rotation and has a temperature close to the virial
temperature of the halo ($T_{\rm halo} \sim 3\times10^6$K for MW type galaxy). This prescription is similar to the 
set-up of \cite{suchkov94}.

To construct the initial density distribution
for the combined gas, we consider the Euler's equation in steady state 
\begin{equation}
-\frac{\nabla p}{\rho} - \nabla \left(\Phi_{\rm DM} +\Phi_{\rm disk} \right) +\frac{v_{\phi,g}^2}{R}\hat{R} = 0 \,,
\end{equation}
for each of the components. Here, $p$ is pressure, $\rho$ is density and $v_{\phi,g}$ is the gas rotation velocity. 
Since for the gas in the disk , the gravitational
force is balanced by rotation and gas pressure together, the rotation speed is less than the particle rotation velocity,
$v_{\phi,G} = \sqrt{R\left[\frac{\partial \Phi}{\partial R}\right]_{z=0}}$, which is solely 
determined by gravity.  For simplicity, we take the rotation velocity of the 
gas as a fraction ($f$ = constant, chosen to be 0.95) of $v_{\phi, G}$ at that point: $v_{\phi,g}=f v_{\phi, G}$.
Therefore, the density distribution for the warm disk  gas can be written as,
\begin{eqnarray}
 \rho_{d}(R,z) &= & \rho_{c}(0,0)\,\exp \Bigl (-\frac{1}{c^2_{\rm sc}} [ \Phi(R,z)-\Phi(0,0) \nonumber\\
 && - f^2(\Phi(R,0)-\Phi(0,0))  ] \Bigr ) \,\,, 
\end{eqnarray}
and for the hot halo gas,
\begin{equation}
 \rho_{h}(R,z) = \rho_{h}(0,0)\, \exp\left(-\frac{1}{c^2_{\rm sh}}\left[ \Phi(R,z)-\Phi(0,0)\right] \right)\,,
\end{equation}
where, $\rho_{c}(0,0)$ and $\rho_{h}(0,0)$ are the warm and hot central densities and 
$c_{\rm sc}$ and $c_{\rm sh}$ are the isothermal sound speeds of the warm disk  and hot halo, respectively (more details on the set-up
are given in appendix \ref{densitysetup}). In Figure \ref{gas_density} we  show the steady state gas density distribution
along the minor axis (red solid line)
and the major axis (black  dotted line). This figure shows that each profile consists of two features, one high density structure
at lower radii representing the disk  material, and another, low density and comparatively flatter distribution at larger radii representing
the halo gas.

In disk  galaxies, along with the thermal pressure,
there is pressure due to turbulence, magnetic fields and cosmic rays, which arises because of the continuous stirring of gas by supernovae.
Therefore, the dynamics of the disk  is determined by both thermal and non-thermal pressures, for which the effective 
sound speed can be written as 
 \begin{equation}
  c_{s,{\rm eff}}^2 = c_{s,T}^2 + c_{s,\sigma}^2 \,,
 \end{equation}
 where, $c_{s,T}$ is the isothermal sound speed corresponding to a temperature $T$ and $c_{s,\sigma}$ is the effective sound speed due to 
 non-thermal components. We include these effects by assuming a disk temperature of $ 4 \times 10^4$ K, with an effective sound speed
 $c_{s,{\rm eff}}=24$ km s$^{-1}$ (larger than $c_{s,T}$, the sound speed of WIM at $10^4$ K).

 For the disk , we set the central density to be $3.0$ particles per cm$^{3}$. 
 The hot halo is however less 
 constrained by observations. In order to fix the central density of the hot halo, we normalise the halo mass distribution
 to give a total halo gas mass $M_{h,{\rm gas}} = 0.11  M_{\rm vir}$ and 
ratio of the stellar disk  mass to the virial mass $M_{\rm vir}$  is $0.05$, as in the scenario of \cite{momao98}.
  Thus,  the halo has a global baryon fraction of $0.16$, consistent with the cosmic value of $f_b = \Omega_b/\Omega_m$.
   Some of the recent observations have revealed that the baryon fraction can be $\sim 0.1$ for massive spirals \citep{bogdan13a}, 
 and in our MW, this fraction can be $\sim 0.16$ for a gas with polytropic index $\gamma=5/3$ in hydrostatic equilibrium \citep{fang13, gatto13}. 
 Though here we assume 
 $f_b = 0.16$, we have also checked the effect of $f_b$ on the mass loading factor at virial radius (see \S \ref{sec:observed_mlf}). 
 
The total density is the sum of the densities of the hot halo and warm disk components, $\rho= \rho_d + \rho_h$. Since the halo gas does not
rotate, the effective rotation speed $v_{\phi,{\rm net}}$ for the 
combined gas is given as
 \begin{equation}
 \rho v_{\phi, {\rm net}}^2= \rho_d \, v_{\phi,g}^2 \,, \quad \Rightarrow \quad
  v_{\phi,{\rm net}} = f\,\sqrt{\frac{\rho_{d}}{\rho} R \left[ \frac{\partial \Phi}{\partial R}\right]_{z=0}} \,.
 \end{equation}

We have found that the above prescription for the initial set up is remarkably stable over
a time scale of 1 Gyr. In reality (in 3D), the interaction between the non-rotating halo gas and the
rotating disk gas could generate instabilities. We can estimate the time scale for Kelvin-Helmholtz
instability at the interface of the rotating disk and the non-rotating halo.  
The dominant wavelength of perturbation is $\sim 10$ kpc,
the corresponding relative velocity $\sim 100$ km s$^{-1}$, and the ratio of densities of two gases
is $\sim 100$ (for gases with temperature $10^4$ and $10^6$ K and in pressure equilibrium). 
The time scale for the growth of perturbations is therefore $\sim 1$ Gyr.
We have also checked with a 3D simulation run that this steady state holds up to $\sim 1$ Gyr.
Therefore the set up described above is adequate for our simulations.

\section{Simulation set-up}
\label{sec:sim_setup}
In this section we describe various simulations that we carry out and the numerical setup. We have carried out
two kinds of simulations to study galactic outflows: small-scale, short-duration (50 Myr) simulations going out to 30 kpc to focus on the inner 
regions where most observational constraints come from; and large-scale, longer-duration (1 Gyr) simulations going out to 250 kpc to study 
cosmological impact of galactic outflows.  
In cases where we focus on observations of multiphase
outflows we use the high-resolution small-scale runs (c.f. Figs. \ref{contour0}, \ref{contour41}, \ref{MT_plot}, \ref{RV}, \ref{tracks}, 
\& \ref{EM}). The large-scale runs are used to infer outflow properties at the halo scale (c.f. Figs. \ref{contour2}, \ref{evolving_rate1}, 
\ref{evolving_rate2}, \ref{velocity_snaps}, \& \ref{vel_mach}).

 We have studied mechanical luminosities ranging from $10^{40.3}$ to $10^{43}$ erg s$^{-1}$ keeping the model parameters fixed.
We have chosen a fiducial run corresponding to a mechanical luminosity injection of 
$\mathcal{L} = 10^{42}$ erg~s$^{-1}$ or SFR $= 14.3$ M$_\odot$~yr$^{-1}$ (see \S \ref{IP}), lying in the intermediate regime of 
luminosities that we have explored.  A full list of runs is given in Table \ref{table:list_runs}.
\subsection{Injection parameters}
\label{IP}
In this paper we only focus on supernovae (SNe) driven outflows. Since a single supernova, or even a large 
number of SNe, is not energetic enough to launch an outflow on larger scales unless they are coherent in space 
and time \citep{nath13,vasiliev14}, we consider the effect of multiple SNe from a large OB association in the 
central region of the galaxy. We consider a constant energy input of mechanical
luminosity $\mathcal{L}$ from the SNe confined in a spherical region of radius $r_{\rm inj}$ at the 
centre of the galaxy. We deposit thermal energy to the gas within $r_{\rm inj}$. \cite{psharma14} have shown
that for such an implementation to work, the injection radius ($r_{\rm inj}$) should be 
such that the energy deposition time is shorter than the cooling time, and we have adjusted our $r_{\rm inj}$ ($= 60$ pc) 
according to this constraint. 
 We also assume that each SN releases an energy of $10^{51}$ ergs, and for a Salpeter mass function, on average,
 $\dot{M}_{\rm inj}=0.1\times$ SFR of mass is injected into the interstellar medium (ISM).
 Therefore, the relation between the mechanical luminosity and SFR can be written as
 \begin{equation}
  \mathcal{L} = 10^{51}\times \, \epsilon \times  f_{\rm SN}\times {\rm SFR}~{\rm erg~s}^{-1} \,,
 \end{equation}
 where, $f_{\rm SN}$ is the number of supernovae explosions per unit mass of stars formed, and
 $\epsilon$ is the efficiency of heating the gas. We assume $\epsilon=0.3$, consistent with
 observations \citep{strickland07} and theoretical estimates from numerical simulations \citep{vasiliev14}. For Salpeter IMF, $f_{\rm SN}=7.4 \times
 10^{-3}\,/M_{\odot}$ for lower and upper limits of stars at $0.1$ and $100$ M$_\odot$. This gives,
 \begin{equation}
 \mathcal{L}= 7\times 10^{40} \, {\rm erg \,\, s}^{-1}\, \frac{\rm SFR}{\left( 1\,{\rm M}_{\odot}~{\rm yr}^{-1} \right)}  \,.
 \label{mdotL1}
 \end{equation}
 This in turn gives the relation between luminosity ($\mathcal{L}$) and the  rate of mass injection as
\begin{equation}
 \dot{M}_{\rm inj} = 0.014\times\frac{\mathcal{L}}{10^{40} {\rm erg}~{\rm s}^{-1}}\,\, {\rm M}_{\odot}\,{\rm yr}^{-1}\,.
 \label{mdotL2}
\end{equation}
The duration of mass and energy injection is assumed to be $50$ Myr, the typical lifetime of an OB association.  The effect of different star burst duration is
are also discussed in \S \ref{sec:multibursts}.

 \begin{table}
  \centering
  \begin{tabular}{ l c l c l} 
   \hline\hline 
  Name & $r_{\rm max}$              & $\mathcal{L}$ & $t_{\rm inj}$ & cooling\\
       & (kpc) & (erg s$^{-1}$)  & (Myr)           &        \\[0.5ex]
   \hline
 L1 & 250 & $10^{40.3}$ & 50 & on \\
 L2 & 250 & $10^{41}$  & 50 & on \\
 L3$^{\ast}$ & 250 & $10^{42}$  & 50 & on \\
 L4 & 250 & $10^{43}$  & 50 & on \\
 L5 & 350 & $10^{41.3}$  & RSB & on \\
 L6 & 250 & $10^{41.7}$  & 50 & on \\
 L7 & 250 & $10^{42.3}$  & 50 & on \\
 L8 & 250 & $10^{42.6}$  & 50 & on \\
 L9 & 250 & $10^{42}$  & 25 & on \\
 L10 & 250 & $10^{42}$  & 100 & on \\
 L11 & 250 & $10^{42}$  & 200 & on \\
 L12 & 250 & $10^{42}$  & 25 & off \\
 L13 & 250 & $10^{42}$  & 50 & off \\
 L14 & 250 & $10^{42}$  & 100 & off \\
 L15 & 250 & $10^{42}$  & 200 & off \\
 S1 & 30 & $10^{40.3}$ & 50 & on \\
 S2 & 30 & $10^{41}$ & 50 & on \\
 S3$^{\ast}$ & 30 & $10^{42}$ & 50 & on \\
 S4 & 30 & $10^{43}$ & 50 & on \\
    \hline
  \end{tabular}
     \caption{ List of runs: The L-series and S-series represents the large-scale and small-scale simulations respectively. The corresponding box size 
     is given in 2nd column, where $r_{\rm max}$ gives the maximum extent of grid in radial direction. The 3rd and 4th column provides the mechanical 
     luminosity and the injection time of the runs, while the 5th column gives the 
     information about cooling. For L5, RSB means Repeated Star Bursts. The fiducial runs are denoted by an ``$\ast$''. Other than these runs, we have also
     run some simulations with variable baryon fraction ($f_b$).}
 \label{table:list_runs}
 \end{table}
\subsection{The code settings}
\label{code}
We use the publicly available hydrodynamic code PLUTO \citep{mignone07} for our simulations. We run our simulations in 2D 
($r,\theta$) spherical coordinates, assuming axisymmetry ($\partial/\partial \phi = 0$). However, we do allow for a non-zero 
azimuthal velocity $v_\phi$.
To solve the hydrodynamic equations (Euler equations with numerical dissipation and with mass and energy source terms which 
drive the outflows), we use piecewise parabolic reconstruction of the primitive variables. We use the advection upstream splitting method 
(AUSM+; \citealt{liou96}) as the Riemann solver and a third-order Runge-Kutta scheme (RK3) to advance the solution in time.

\begin{figure*}
 \includegraphics[trim=2.0cm 1.0cm 1.0cm 0.0cm, clip=true, totalheight=0.67\textheight, angle=-90 ]{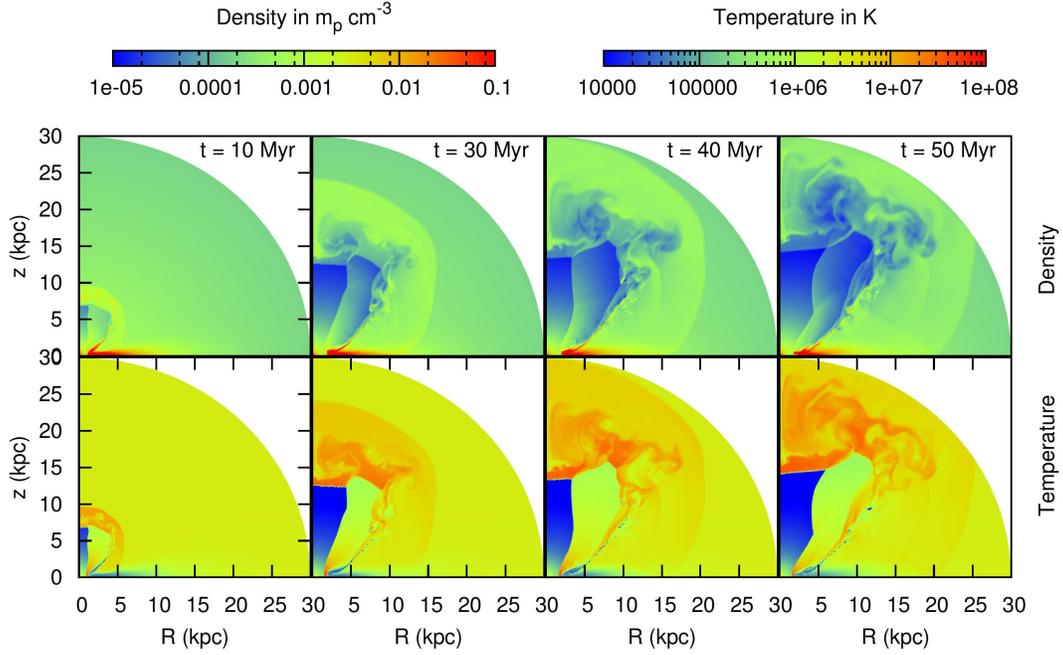}
 \caption {Snapshots of density (top panel) and temperature (bottom panel) at 10, 30, 40 and 50 Myr  for 
           $\mathcal{L} = 10^{42}$ erg s$^{-1}$ for a box size of $r_{\rm max} = 30$ kpc. 
           Notice that the cold, multiphase gas, which is mainly due to the uplifted disc gas, is confined to the outer wall of the outflow. 
           }
 \label{contour0}
\end{figure*}

\begin{figure*}
 \includegraphics[trim=2.0cm 1.0cm 1.0cm 0.0cm, clip=true, totalheight=0.67\textheight, angle=-90 ]{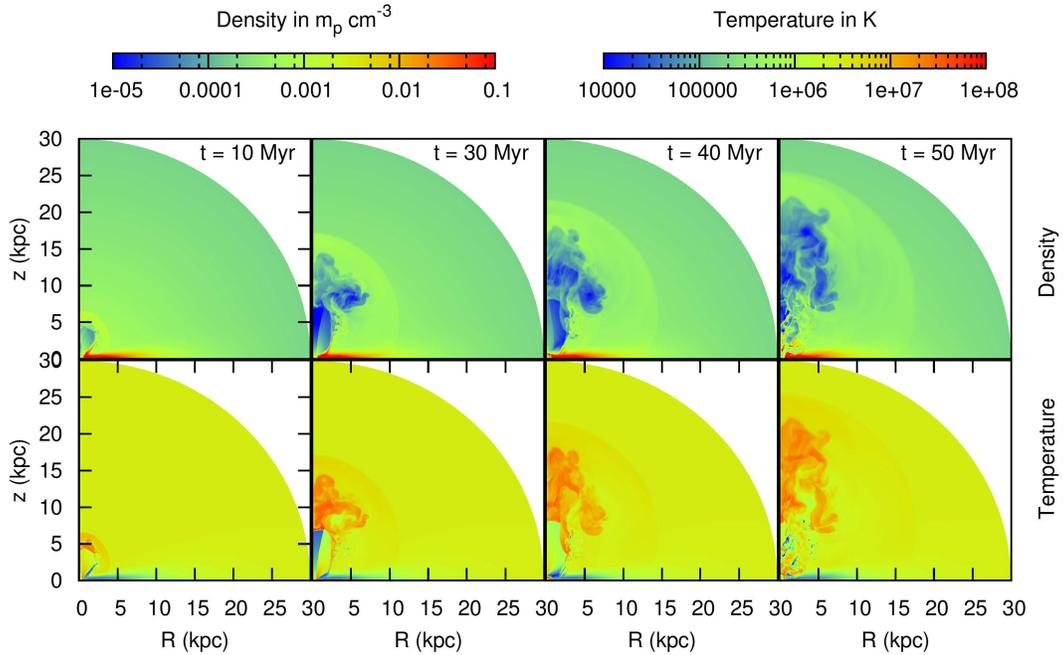}
 \caption {Snapshots of density (top panel) and temperature (bottom panel) at 10, 30, 40 and 50 Myr for $\mathcal{L}=10^{41}$ erg s$^{-1}$ for a box 
          size of  $r_{\rm max} = 30$ kpc.
          The evolution is different from Figure \ref{contour0} in that, in addition to the cold gas at the outer wall, 
          there is volume-filling cold disk gas at 50 Myr dredged up by the ram pressure of the outflow.
          It will be shown later in \S \ref{VS} that the cold gas at the outer wall is slower compared to the volume-filling cold gas.}
 \label{contour41}
\end{figure*}

\begin{itemize}
\item \textit{Grid :} Since we inject the SNe energy and mass in a small region of space 
($= 60$ pc) at the centre of the galaxy and try to observe the effects near the virial radius ($\sim 250$ kpc),
we use a 
logarithmic grid in the radial ($r$) direction. It starts from $20$pc and goes till $250$ kpc for the large scale simulations 
and till 30 kpc for the small-scale simulations. A uniform grid is used in the $\theta$ direction going from $\theta = 0$ to $\pi/2$.
\footnote{More can be found in the user's guide of PLUTO - http://plutocode.ph.unito.it/Documentation.html}. 
For the results mentioned in this paper, we use 512 grid points along both $r$ and $\theta$ directions. We have also
carried out resolution studies with double and half this resolution in each direction.
\item \textit{Boundary conditions :} The
inner and outer radial boundary values for  mass and energy densities are set to
their equilibrium values at $t = 0$ (as discussed in \S\ref{HE}). The velocities are
copied in the radial ghost zones from the nearest active zones.
The $\theta$ boundary conditions are set as reflective.
\item \textit{Metallicity:} Since the mixing of metals at kpc range in ISM densities and temperatures 
is dominated by the dynamical evolution of the gas rather than diffusion, we track the metallicity by treating
it as a passive scaler which follows the simple advection equation. We set the disk  metallicity to be equal to the 
solar metallicity ($\mathcal{Z}_{\odot}$) and the halo metallicity to be $0.1\,\mathcal{Z}_{\odot}$.
\item \textit{Cooling :} PLUTO can include optically thin losses in a fractional step formalism$^1$. It 
has several different cooling modules, among which, we use the tabulated cooling method which solves the 
internal energy equation from a given $T-\Lambda(T)$ table.
We include the metallicity effect in the cooling rate by using linear interpolation of the cooling curves corresponding 
to $\mathcal{Z} = \mathcal{Z}_{\odot}$ and $\mathcal{Z} = 0.1\mathcal{Z}_{\odot}$ \citep{sutherland93},
 to all other metallicities from $\mathcal{Z}_{\odot}$ to $0.1 \mathcal{Z}_{\odot}$ . 
 
In our calculation, we express the temperature as $T = p/\rho$, which 
includes the dynamical pressure in addition to the thermal pressure, and therefore the  effective temperature of 
the gas in the disk becomes large enough to induce strong cooling, unlike in the WIM at $10^4$ K.
 To stop this cooling, we constrain the cooling function of the disk  material (but not the injected material) to be zero
within a box of size $R\times z = 15\times2$ kpc$^2$, 
This can be thought of as a crude model of 
continuous SNe/stellar heating or turbulent support 
of gas in the disk, which prevents disk cooling. 
\item \textit{Units :} To avoid the calculation of very small ($\sim 10^{-24}$) or very large ($\sim 10^{33}$) numbers,
PLUTO works with non-dimensional, arbitrary units. The basic units used in our simulations are length ($L_0$)
$ = 1$ kpc, velocity ($v_0$) $= 100$ km s$^{-1}$ and density ($\rho_0$) $= 1.67\times 10^{-23}$gm cm$^{-3} = 10$ m$_p$ cm$^{-3}$. All 
other units are derived from these basic units as time ($t_0$) $= L_0/v_0\,= 9.8$ Myr and pressure
($p_0$) = $\rho_0 v_0^2 = 1.67\times 10^{-9}$ dyne cm$^{-2}$.
Therefore, the rate of energy and mass injection to the spherical starburst 
region (using Eq. \ref{mdotL1} and  \ref{mdotL2}) can be written in terms of pressure and density as
\begin{align}
\dot{p} \: &= \frac{2}{3}\,\frac{\mathcal{L}}{\left(4\pi/3\right)\,r_{\rm inj}^3} \:\:  
\nonumber\\
        &= 9.7\times \left(\frac{\mathcal{L}}{10^{40} \,{\rm erg}~{\rm s}^{-1}} \right) \left(\frac{100\,{\rm pc}}{r_{\rm inj}} \right)^3\:\: p_0/t_0 \\
\dot{\rho} \: &= \frac{\dot{M}_{\rm inj}}{(4\pi/3) r_{\rm inj}^3} \nonumber \\
        &= 0.118\times \left( \frac{\mathcal{L}}{10^{40} \, {\rm erg}~{\rm s}^{-1}}\right)\left( \frac{100\,{\rm pc}}{r_{\rm inj}}\right) \:\: \rho_{0}/t_0
\end{align} 
where, $p_0$, $\rho_0$ and $t_0$ are the code units of pressure, density and time respectively.
\end{itemize}
\begin{figure*}
  \includegraphics[trim=0.0cm 0.0cm 0cm 0cm, clip=true, height=0.65\textheight, width=0.5\textheight,angle=-90]{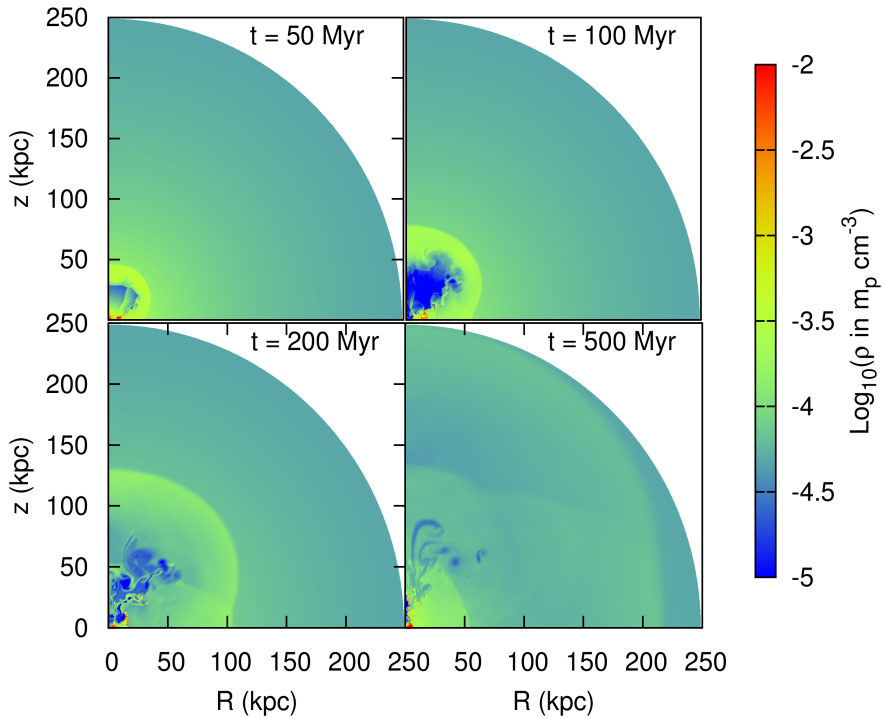}
 \caption {Density distribution at 50, 100, 200 and 500 Myr for $\mathcal{L}=10^{43}$ erg s$^{-1}$
           for a box size of $r_{\rm max} = 250$ kpc. Note that these plots are on a scale $\sim 10$ times larger than 
           the ones in Figures \ref{contour0} \& \ref{contour41}.
   }
 \label{contour2}
\end{figure*}
\section{Results}
In this section we present our simulation results on the effect of outflows with different mechanical luminosities.
Figure \ref{contour0} shows
the evolution of density and temperature for the zoomed-in (box size of $30 \times 30$ kpc$^2$) fiducial run ($\mathcal{L} = 10^{42}$erg s$^{-1}$). 
It shows the standard stellar wind structure, 
with a free wind (\citealt{CC85}) in the inner region characterised by dilute gas with high velocity, surrounded by the 
shocked wind and the shocked ISM. After breaking out from the disk, the free wind forms a conical shape,
because of the interaction with the halo gas. The shocked wind and the shocked ISM form a multiphase structure.
Because of radiative cooling, 
parts of this interaction zone with high density gas breaks into smaller clumps and forms clouds which are then carried away by the outflow 
or they fall back towards the galactic center due to gravity.

Similar wind structure is found in general for all mechanical luminosities. However, for low luminosity
cases, because of the relatively low pressure in the central region, the disk  material can press inwards after a certain time. 
Figure \ref{contour41}
shows one such example for $\mathcal{L}=10^{41}$ erg s$^{-1}$ 
where, at 40 Myr, the advancement of the disk  material 
almost completely covers the injection region  (at a scale of $\sim 1$ kpc). The increase in pressure, because of the continuing injection of 
energy and mass, thereafter blows away the disk  material and forms a filamentary multiphase outflow.
We discuss
the characteristics of these clouds and filaments later in \S\ref{VS}. The interference of the disk  material into the 
base of free wind can also add ripples to the conical shape of the free wind as witnessed in Figure \ref{contour0}.

After injection is switched off ($t > 50$ Myr), the free wind disappears and the inner region develops a complex density and velocity 
structure, which then gradually falls back towards the galactic center as shown in Figure \ref{contour2}.
The forward shock, however, keeps propagating through the halo medium, finally becoming an acoustic disturbance
of the medium for lower luminosities ($L \lesssim 10^{42}$ erg s$^{-1}$). 
The material that falls back to the center with non-zero velocity, collides with other gas clumps and generates secondary shocks which
then push the infalling material away (lower panel of Fig. \ref{contour2}). After few such bouncing back and forth, the gas finally settles
down at the center to form a disk-like structure (right-bottom panel of Fig. \ref{contour2}). In this 
whole process of infall and outflow, some dilute ($\rho \sim 10^{-5}$ m$_p$cm$^{-3}$) and hot ($T\sim 10^{7}$K) gas is left behind
in the halo in the form of eddies. This gas neither takes part in outflow nor contributes to infall, rather becomes a part of the circumgalactic
medium via dynamical mixing.

Quantitatively, the parameters of interest are, 
1) the mass loading factor, which tells us the amount of mass that goes out of the virial radius,
2) the temperature distribution of the outflowing material, which holds the information regarding the various phases of
the gas, and which determines the observability of these phases.
3) the velocity structure, which gives an overview of the motions of different gas phases and the coupling between them.
We will discuss these properties, below, one by one.

\subsection{Mass loading factor}
\label{sec:mlf}
The mass outflow rate across a spherical shell of radius $r$ can be written as
\begin{equation}
\dot{M}_{\rm out}\left(r,t\right) = 4 \pi r^2 \int_{0}^{\pi/2} (\rho v_r) \sin\theta\, d\theta \, ,
\end{equation} 
where $\theta$ is the zenith angle. Note that the velocity $v_r$  in the integrand has both positive and negative values, so that, 
after integrating, it gives the net mass outflow rate. This outflow rate is finally  integrated over time to obtain the 
 total amount of outflowing mass ($M_{\rm out}$) at each radius for different luminosities. 

This mass outflow can be compared with the 
 total mass of new stars formed ($M_{\ast}^{+}$), and one defines a mass loading factor as
\begin{equation}
\eta(r,T) = \frac{M_{\rm out}}{M_{\ast}^{+}} = \frac{1}{M_{\ast}^{+}}\, \int_0^T \dot{M}_{\rm out}(r,t)\,dt\,.
\label{eq:eta_def}
\end{equation}
 The choice of the integration time, $T$, depends upon the spatial scale of interest and is discussed later in this section.

\begin{figure}
 \includegraphics[width=6.0cm,angle=-90 ]{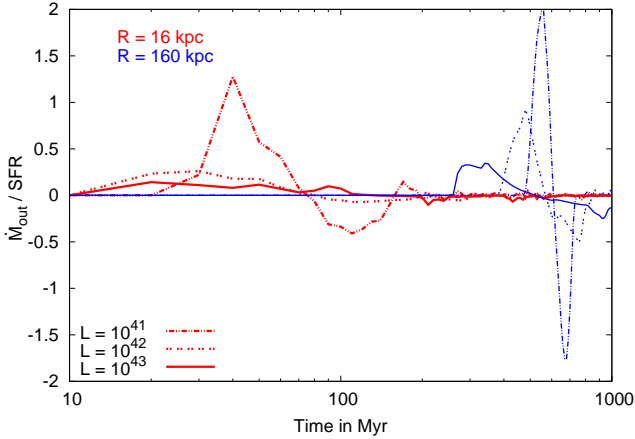}
 \caption {The evolution of the mass outflow rate in the units of starburst SFR 
 at two different radii, $16$ kpc (red,  thicker lines) and
 $160$ kpc (blue,  thinner lines) as a function of time (in Myr) for different luminosities
 (shown in different line styles: $10^{41}$ erg s$^{-1}$ (dot-dashed), $10^{42}$ erg s$^{-1}$ (double dotted) and $10^{43}$ erg s$^{-1}$ (solid). 
 The results are for a simulation box size of $r_{\rm max} = 250$ kpc.
   }
 \label{evolving_rate1}
\end{figure}

Figure \ref{evolving_rate1} shows the evolution of the mass  outflow rate $\dot{M}_{\rm out}$ in units of starburst SFR
at two different radii, $ r = 16, 160$ kpc  (shown with red (thick) and blue (thin) lines),  as a function of time (in Myr) for three different injection
luminosities, $10^{41, 42, 43}$ erg s$^{-1}$ (shown with dot-dashed, double dotted and solid lines, respectively). 
Consider the blue and red dot-dashed lines, denoting the evolution of $\dot{M}_{\rm out}$ for $L=10^{41}$ erg s$^{-1}$
with time. We find that a shell of shocked ISM and shocked wind travels outward, reaching $\sim 160$ kpc (blue) in $\sim 600 $ Myr.
The negative values of $\dot{M}_{\rm out}$ corresponds to infall, which at the very outer radii arises due to acoustic oscillations of the halo gas, 
but at inner radii corresponds to the infall of gas due to various instabilities. The interaction region between the halo and the wind 
suffers from thermal and Kelvin-Helmholtz instabilities, as well as the Rayleigh-Taylor instability when there is acceleration 
\citep{fraternali06, sharma13b}. These instabilities are the key mechanisms behind the formation of clouds, some of which are the 
part of a galactic fountain.

\begin{figure}
  \includegraphics[width=6.0cm,angle=-90 ]{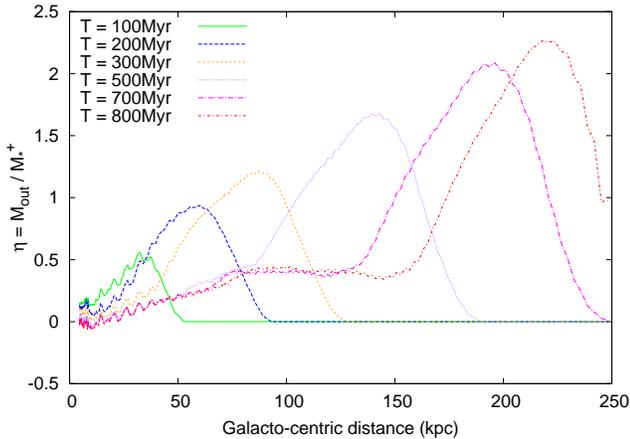}
 \caption { Mass loading factor ($\eta$) as a function of the galacto-centric radius, for $\mathcal{L}=10^{42}$ erg s$^{-1}$.
  The values of $\eta$ shown in different colours/styles correspond to different periods of  integration time, from 100 Myr to 800 Myr. 
           }
 \label{evolving_rate2}
\end{figure}

The time  integrated outflowing mass presents a less chaotic behaviour. In Figure \ref{evolving_rate2} we show 
the time integrated values of  the mass loading factor $\eta$ (eqn \ref{eq:eta_def}) for our fiducial run  ($\mathcal{L}=10^{42}$ erg s$^{-1}$), 
 integrated over different periods for each distance (shown in different colours/styles). 
The figure shows the gradual outward progression of the outflowing material in the halo. E.g., the shell reaches a distance of
 $\sim 50$ kpc in $100$ Myr, and finally reaches the virial radius at a time scale of $\sim 800$ Myr. 
The figure also shows that the mass loading factor in the inner region can have small values when 
 integrated over a long time scale, because of the infall of material in absence of injection.
Therefore the behaviour of $\eta$ at the inner region can be better understood if it is  integrated
over an appropriate, and short time scale ($\sim 100$ Myr). On the contrary, 
for the mass loading factor  near the virial radius, it is reasonable to average it over a long ($\sim$ Gyr)
time scale.
 The mass outflow  at smaller scale is commonly compared with the current (ongoing) star formation process.
However, since the effect of central starburst reaches the virial radius only after sufficient time has elapsed (the travel time, $\sim$ Gyr), 
the mass outflow  at large radii cannot be connected to the present day star formation. Instead, it should be compared with 
the past SF which caused it (i.e. the injection epoch). We define the outer mass loading factor (to be precise, the mass loading factor at virial radius)
as $\eta_v = \eta(r_{\rm vir},t_v)$, where the integration time $t_v$ is the roughly the time taken by outflow 
to reach $200\hbox{--}230$ kpc. Depending on the luminosity, $t_v$ varies from $750$ Myr to $900$ Myr. In order to evaluate the inner mass
loading factor we take integration time $t_{v} = 100$ Myr. Since the shell of outflowing material reaches a distance
$20\hbox{--}50$ kpc in $100$ Myr, we denote the inner mass loading factor as $\eta_{\rm 20}$. 
The outflow is mainly contributed by a shell of mass moving out through the medium (Figure \ref{evolving_rate2}). Therefore, 
we take the peak value of the shell as the mass loading factor at that epoch.

\begin{figure}
 \includegraphics[width=3.0cm,angle=-90 ]{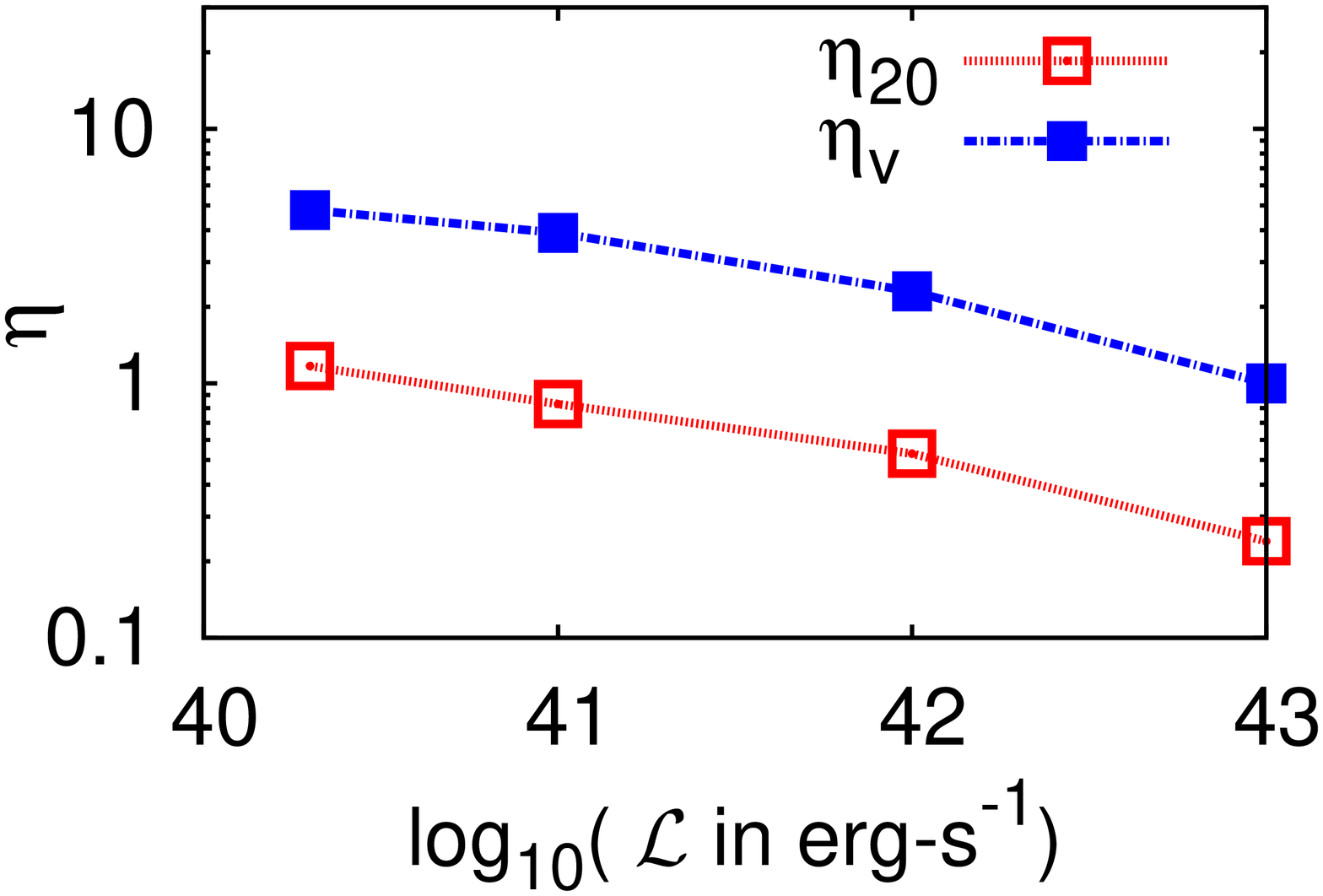}
 \includegraphics[width=3.0cm,angle=-90]{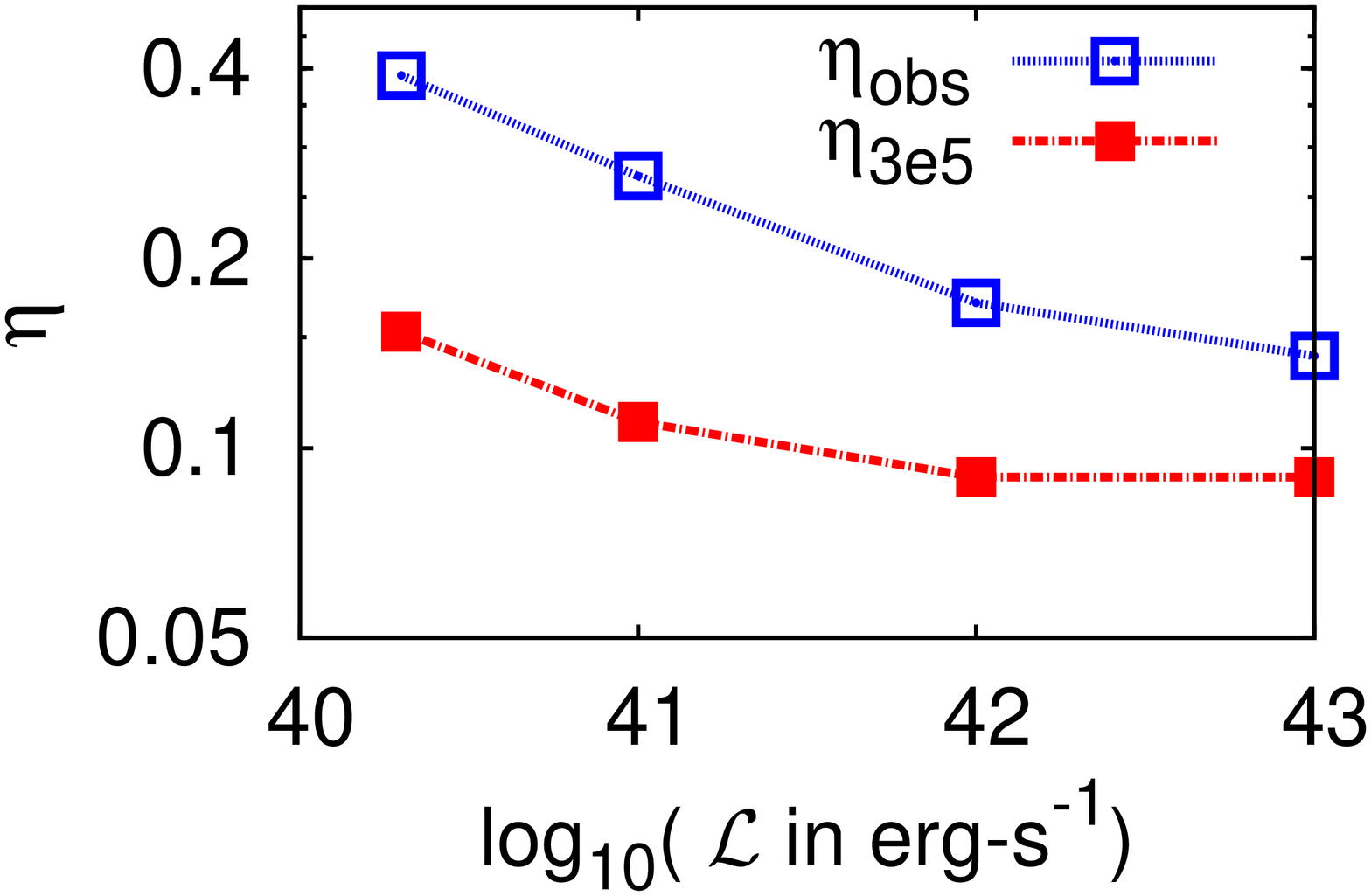}
 \caption { Left panel: The mass loading factor ($\eta$) at smaller radii (red open squares) and at virial
           radius (blue filled points)  based on large-scale runs. Right panel: Observational mass loading factor (Eq. \ref{eta_obs})
           based on small-scale runs. The blue open squares show the total $\eta_{\rm obs}$ and red filled squares show the warm mass loading factor.
           }
 \label{eta_L1}
\end{figure}

The dependence of these two values of $\eta$  (i.e. $\eta_{\rm 20}$ and $\eta_{v}$) on $\mathcal{L}$ is shown in the left
panel of Figure \ref{eta_L1}, where the red open squares
show the values of $\eta_{20}$, appropriate for the inner region, and the blue filled squares show $\eta_v$, the values at
the virial radius. The curves show that in the inner regions, the mass loading factor decreases with $\mathcal{L}$, ranging
between $\sim 0.3 \hbox{--}1$, with an approximate power-law scaling $\eta \propto \mathcal{L}^{-0.25}$. The values at the outer radii 
also scale with $\mathcal{L}$ with a similar power-law index,  and ranges between $\eta \sim 1.0\hbox{--}5.0$.
 The negative slope of $\eta$ can be understood with very simple arguments. Consider a blast wave with energy $E$ propagating in an
uniform density medium. The shock radius and velocity can be given as $r_{os} \sim E^{1/5}$ and $v_{os} \sim E^{1/5}$ at any particular time. 
Therefore the mass outflow rate inside the shell can be written as $\dot{M}_{\rm out} \sim r^2\,v$ which in turn gives $\dot{M}_{\rm out}/E \sim E^{-2/5}$,
which is equivalent to the mass loading factor we have defined here. Hence, the negative dependence of $\eta$ on $\mathcal{L}$ appears naturally.
Physically, explosions with smaller $\mathcal{L}$ produce an outflow that is strongly coupled
to the halo gas, because (a) of low speed and (b) small conical angle in which the outflow is confined. Strong explosions, on the other hand,
tend to propagate through the halo gas quickly, sweeping it with high speed, instead of much coupling.

 We have checked that in the absence of the hot halo gas, the mass-loading factor $\eta$ approaches 0.1 in the whole
range of $\mathcal{L}$  considered here. It is connected with the fact that the typical mechanical luminosities in active central starbursts
are always much 
higher than the critical luminosity necessary to break through the galactic ISM disk : $\mathcal{L}_{\rm cr}\sim 10^{38}$ erg s$^{-1}$ for
the ISM parameters used above \citep{nath13}.  Once the wind breaks out of the disk, there is no halo resistance to stop it.
Therefore, the total outflowing mass is equal to the injected mass, and the mass loading factor becomes  equal to the injection value ($\eta_{\rm inj}$), 0.1 
(see Fig. \ref{eta_fb}).

\subsection{Effect of multiple bursts and injection time}
\label{sec:multibursts}
 So far we have considered the effect of a single starburst (injecting from $t=0$ to $50$ Myr). The real situation may, however, differ in 
particular cases as there may be multiple bursts at the centre, or the injection time period may differ. We consider two extreme cases. 
First, in which the active star formation periods 
are well separated in time from each other, i.e. they are almost independent event. In the second case, the star bursts are so close in time that they can
be considered as a continuous event. We use our fiducial run to compare with the other variants (with different $\mathcal{L}$ and $t_{inj}$).

\begin{figure}
 \includegraphics[width=6.0cm,angle=-90 ]{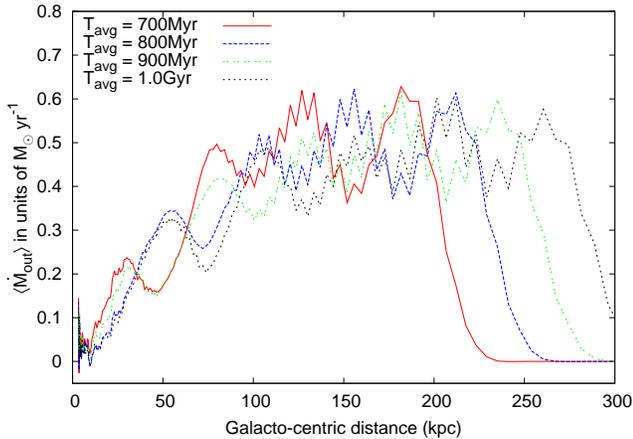}
 \caption { Effect of multiple bursts, each of $\mathcal{L} = 10^{41.3}$ erg s$^{-1}$ and $t_{\rm inj} = 50$ Myr, on the time averaged mass outflow rate,
          $\langle \dot{M}_{\rm out} \rangle$, at each radius. The individual peaks corresponding to the individual
          starburst events show almost similar behaviour. The box size in this case is $r_{\rm max} = 350$ kpc.}
 \label{multibursts}
\end{figure}

 The first case is implemented by putting starbursts of $\mathcal{L} = 10^{41.3}$ erg s$^{-1}$ with $t_{\rm inj} = 50$ Myr at large time separation. 
We put five such 
starbursts at the centre separated by $200$ Myr in time so that the total injected energy becomes equal to the fiducial value ($\mathcal{L} = 10^{42}$ 
erg s$^{-1}$). We have calculated the average mass outflow rate as,
$\langle \dot{M}_{\rm out}(r) \rangle = \frac{1}{T_{\rm avg} }\int_0^{T_{\rm avg}} \dot{M}_{\rm out}\,dt$, where $T_{\rm avg}$ is varied between $700$ Myr and 
1 Gyr. This choice of $T_{\rm avg}$ is motivated by the fact that the shell corresponding to the first burst takes roughly a Gyr to reach the virial radius, 
and subsequent shell lags behind it by roughly $200$ Myr.
The result of such bursts is shown in Fig. \ref{multibursts}. The figure clearly shows that the individual outflowing shells
corresponding to the independent starbursts move through the halo medium almost uninterrupted by the previous bursts and create the same 
effect as it would have done for a single burst of $\mathcal{L} = 10^{41.3}$ erg s$^{-1}$. Hence, multiple starbursts separated by a long time interval
can be treated in the same way as we treat an individual burst.
\begin{figure}
 \includegraphics[width=6.0cm,angle=-90]{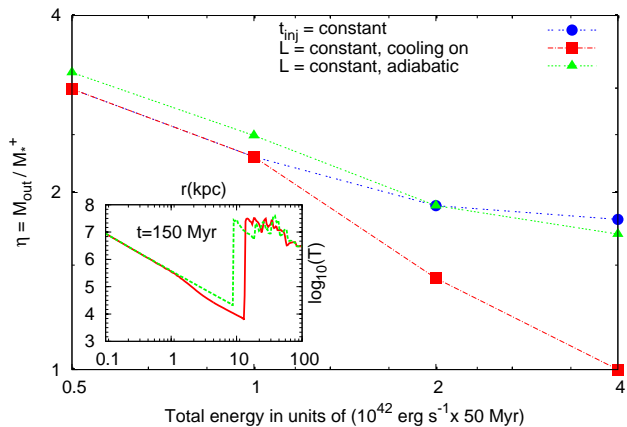}
 \caption{ The effect of increasing energy on the mass loading factor $\eta_v$ is shown by increasing the injection time and injection mechanical
         luminosity. The blue dots 
         show the results when the injection time is constant ($t_{inj} = 50$ Myr) and luminosity is varied. The red squares show the result when the 
          mechanical luminosity constant ($\mathcal{L} = 10^{42}$ erg s$^{-1}$)
         but energy is increased by increasing the injection time. The green triangles show the run for variable $t_{\rm inj}$, but, with cooling
         switched off. The inset shows the temperature profiles at $t = 150$ Myr for $t_{\rm inj} = 200$ Myr run. The red (solid) line corresponds 
         to the case where radiative cooling is switched on, and the green (dotted) line represents the case where radiative cooling is absent.}
   \label{effect_of_tinj}
\end{figure}

 In order to understand the effect of $t_{\rm inj}$ on $\eta_v$, we use $\mathcal{L} = 10^{42}$ erg s$^{-1}$ with different 
injection times ($t_{\rm inj}$) of $25,\,100,\,$ and $200$ Myr. The increment in the injected energy due to the increased $t_{\rm inj}$ gives 
rise to lower mass loading factor compared to the constant $t_{\rm inj}$ cases. The values of $\eta_v$ for these cases  depend on the total energy as
$\eta_v \propto \mathcal{E}^{-0.5}$ (see Fig. \ref{effect_of_tinj}), where $\mathcal{E}=\mathcal{L} \times t_{\rm inj}$.
Figure \ref{effect_of_tinj} shows that $\eta_v$ decreases when the injection time lasts for more than $\sim 50$ Myr compared to
runs which have same energy but where the injection lasts only for $50$ Myr.

 A comparison with the adiabatic counterparts of these runs (shown by the 
green triangles in Fig. \ref{effect_of_tinj}) shows that this decrease in the mass loading factor can be attributed to the radiative cooling of the 
free wind which lasts long enough ($> 50$ Myr) to radiate away a significant fraction of the total energy. The inset of Fig. \ref{effect_of_tinj} shows 
the temperature profiles of a ($10^{42},\,200$ Myr) run at $t = 150$ Myr. This plot shows the decrease of temperature (or the internal energy)
due to radiative losses in the free wind. Thus it is evident that if the free wind phase lasts for a long time (due to prolonged injection time), then
it radiates away a good fraction of the energy.

 The above results allow us to write the variation of $\eta_{\rm 20}$ and $\eta_v$ as a function of the total energy or total mass of stars formed 
within the injection time (obtained from the fits of Fig. \ref{eta_L1}) as
\begin{equation}
\eta_{20} \approx 0.4\times \left( \frac{\mathcal{E}}{\mathcal{E}_F} \right)^{-0.25} = 0.4\times \left( \frac{M_{\ast}^{+}}{M_{\ast}^F} \right)^{-0.25},
\label{eq:eta_20}
\end{equation}
and
\begin{equation}
\eta_{v} \approx 2.5\times \left( \frac{\mathcal{E}}{\mathcal{E}_F}\right)^{-0.25} = 2.5\times \left( \frac{M_{\ast}^{+}}{M_{\ast}^F}\right)^{-0.25}\,.
\label{eq:eta_v}
\end{equation}
 where, $\mathcal{E}_F = 10^{42}$ erg s$^{-1} \times 50$ Myr  is the energy for the fiducial run and $M_{\ast}^F (= 14.0$ M$_{\odot}$ yr$^{-1} \times 50$
Myr) is the corresponding mass of new stars formed. However, for long $t_{\rm inj}$ ($\gtrsim 50$ Myr) in case of $\mathcal{L} > 10^{42}$ erg s$^{-1}$, 
the cooling affects the dynamics and the mass loading factor can be written as
\begin{equation}
\eta_v \approx 2.5\times \left( \frac{\mathcal{E}}{\mathcal{E}_F}\right)^{-0.5} = 2.5\times \left( \frac{M_{\ast}^{+}}{M_{\ast}^F}\right)^{-0.5}\,.
\end{equation}
\subsection{Comparison with observed mass loading factor}
\label{sec:observed_mlf}
We compare the values of $\eta$ obtained from our simulations with those estimated from observations, where only partial information about 
the velocity and density structure is available. The mass loading factor in observations is defined as $\eta
= \frac{M\times v_r}{r\times {\rm SFR}}$, where, $M$ is the total mass of outflowing gas (observed as molecular or ionised gas), $r$ is the typical
scale of the outflowing region and $v_r$ is an estimate of the outflowing gas velocity \citep{arribas14, bolatto2013} or the sound speed \citep{strickland07}. 
In order to compare with the observed values of mass loading factor, we define $\eta_{\rm obs}$, the mass loading factor within a outflowing region 
of radius $r_{d}$ at any time $t$ as
\begin{equation}
\label{eta_obs}
\eta_{\rm obs}(r_{d},t) = \frac{ M\,v_r }{{\rm SFR}\,\,r_d} = \frac{1}{\rm SFR}\frac{2\pi}{r_d} \int_0^{r_d} r^2\,dr\,\int_0^\pi (\rho v_r) \, \sin \theta \, d\theta \,,
\end{equation}
where, $r_d$ is taken to be $10$ kpc, the radius within which most of the observations are limited. It is to be noted that the velocity inside the integral
is the actual velocity of any individual fluid packet rather than some characteristic speed of the whole fluid as usually considered by the observations.
Since, during a starburst the mass outflow rate is not constant because of the halo-wind interactions
and formation of clouds and eddies, to get a reliable value, and to connect with the current SFR, 
we average it over the injection period ($50$ Myr). The values of time averaged $\eta_{\rm obs}$
is shown in the right panel of Figure \ref{eta_L1} by the  blue open squares. These values show the same behaviour as seen previously
in $\eta_{20}$ only with a shallower dependence on $\mathcal{L}$ ($\eta_{\rm obs} \propto \mathcal{L}^{-0.15}$).  

\begin{figure}
 \includegraphics[trim=3cm 3cm 0cm 0cm, clip=true, totalheight=0.4\textheight, angle=-90]{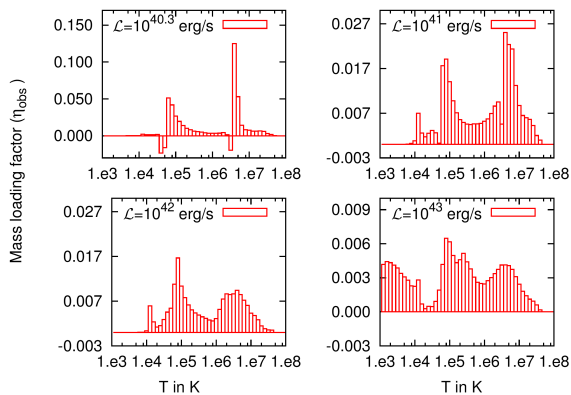}
 \caption{Temperature distribution of the outflowing gas for different luminosities at a galacto-centric radius $10\,$kpc,
          averaged over $50\,$Myr. }
 \label{MT_plot}
\end{figure}

However, the estimation of the mass loading factor from observations is either for the cold molecular gas 
\citep{bolatto2013} or the ionised \citep{arribas14} or the hot gas \citep{strickland07}, and not for all the phases taken together. Therefore, to determine 
 the mass loading factors for different phases and find the correlation between them,
 it is important to study the temperature distribution
of $\eta_{\rm obs}$.
Figure \ref{MT_plot} shows that the outflowing mass is divided
mainly into two temperature domains, one at $\sim 10^5$ K, and, another at $\sim 5\times 10^6$ K. The $10^5$ K gas comes from the evaporation of the 
disk gas and adiabatically expanded wind material, and, the hot gas (T $\sim 5\times10^6$ K) comes from the shocked ISM and wind material.
We also notice a small peak near 
$T = 10^4$K, which arises because of the clouds formed from the interaction between wind and halo material and from the condensation of the 
evaporated disk  material. The extended
outflowing gas at $T < 10^4$ K in case of $\mathcal{L} = 10^{43}$ erg s$^{-1}$ arises due to the
adiabatic cooling of the free wind (see right-bottom plot of Figure \ref{MT_plot}).

The temperature plots imply that observations aimed to detect either the cold or ionised or hot gas are likely to miss a significant
fraction of the outflowing material even at small radii. Therefore, to determine the contribution of the warm gas which is 
the commonly used observational tracer of outflow, we define a new mass loading 
factor $\eta_{\rm 3e5}$ which counts only $T < 3\times10^5$ K gas. This
is represented by red filled squares in the right panel of Figure \ref{eta_L1}, which also shows that $\eta_{\rm 3e5}$ is less than the total mass loading factor 
$\eta_{\rm obs}$ by a factor of $2\hbox{--}3$. Moreover, the mass loading factor is almost equal to the injection value, $0.1$ (see \S\ref{IP}), 
which is of a similar magnitude as estimated by \cite{arribas14} in case of MW type galaxies (dynamical mass $\sim 10^{11}M_{\odot}$).
 
 To understand the relations between different mass loading factors we have fitted them with simple power-law relations:
\begin{eqnarray}
 &\eta_{\rm obs}& \simeq 0.4\times \mathcal{L}_{40}^{-0.15} \simeq 0.3\times \left( \frac{\rm SFR}{{\rm M}_\odot~{\rm yr}^{-1}} \right)^{-0.15} \,,\\
 \label{eta4}
 &\eta_{\rm 3e5}& \simeq 0.15 \times \mathcal{L}_{40}^{-0.1} \simeq 0.12\times \left( \frac{\rm SFR}{{\rm M}_\odot~{\rm yr}^{-1}} \right)^{-0.1}\,.
 \end{eqnarray}
 Comparing these equations with Eq. \ref{eq:eta_v} and using Eq. \ref{mdotL1}, we can write
\begin{eqnarray}
\label{etarelations2}
\eta_{v} &\approx & 5 \times \left( \frac{\rm SFR}{{\rm M}_{\odot} {\rm yr}^{-1}}\right)^{-0.25}\times \left(\frac{t_{\rm inj}}{50 {\rm Myr}} \right)^{-0.25}
\nonumber\\
&\approx& 40 \times \left( \frac{\rm SFR}{{\rm M}_{\odot} {\rm yr}^{-1}}\right)^{-0.15}\times \left(\frac{t_{\rm inj}}{50 {\rm Myr}} \right)^{-0.25}
\times \eta_{\rm 3e5}\,. 
\end{eqnarray}
This gives a relation between the warm mass loading factor, $\eta_{\rm 3e5}$ (relevant for observations), and the outer mass loading factor ($\eta_v$) (relevant for
 cosmological scales)  for a given starburst period and SFR. Eq. \ref{etarelations2} shows that the mass loading factor at the virial radius
is larger by a factor $\sim 40$ than the mass loading factor that is observable near the central region. 
This relation is almost independent of SFR but depends upon the starburst activity time.

\begin{figure}
 \includegraphics[width=6.0cm,angle=-90 ]{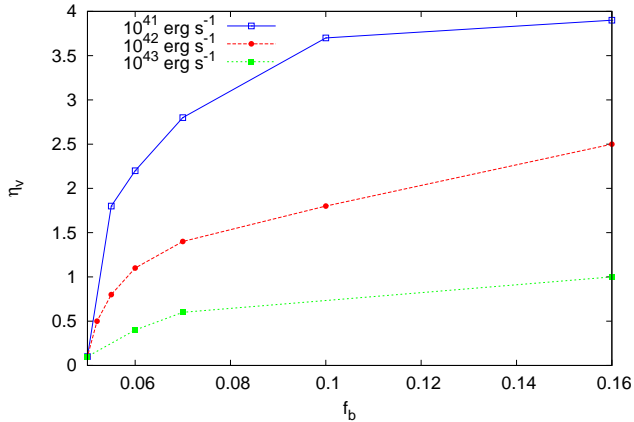}
 \caption { Variation of outer mass loading factor ($\eta_v$) with baryon fraction of the galaxy in the case of different luminosities of 
          $\mathcal{L} = 10^{41, 42, 43}$ erg s$^{-1}$ (or runs L1, L2 and L3 in Table \ref{table:list_runs}).}
 \label{eta_fb}
\end{figure}

 The ratio  $\eta_v / \eta_{\rm 3e5}$ also depends on the baryon fraction of the galaxy, in particular, the fraction of the total
mass that is in the form of halo gas. Since the main contribution to the outflowing mass at
the outer radii comes from the swept up halo material, the outer mass loading factor strongly depends on the mass budget of the
background halo gas. Figure \ref{eta_fb} 
shows the variation  of $\eta_v$ as a function of the baryon fraction ($f_b$). The figure shows that, for $f_b \gtrsim 0.1$ i.e. when more than half of the 
baryon is in the hot halo phase, $\eta_v$ varies weakly with $f_b$. However, for $f_b \lesssim 0.1$, mass loading factor decreases steeply
and finally for $f_b = 0.05$ (the stellar mass fraction), it becomes equal to the injection value, $0.1$. Therefore, $\eta_v / \eta_{\rm 3e5}$ can vary
between $\sim 1\hbox{--}40$ depending on the baryon fraction.

 In this paper, we have only varied the mechanical luminosity and injection time; the other parameters like total galactic mass have been kept fixed.
The observational scaling of $\eta(\mathcal{\rm SFR})$, on the other hand, may be contaminated by these additional variables. Indeed, a larger mechanical 
luminosity suggests, in general, a higher star formation rate, which in turn may indicate a larger galactic mass, and a more massive and extended hot halo.

We have also carried out simulations at various resolutions, and we find excellent convergence for the various mass loading factors 
($\eta_{20}$ \& $\eta_v$) in case of the large-scale runs. The small-scale runs, which study the multiphase gas, are somewhat resolution-dependent as they do 
not resolve the transition regions between the cold and hot phases in the absence of thermal conduction \cite[see][]{koyama04}. However, the relation between
the mass loading factors in Eq. \ref{etarelations2} holds within a factor of two for all resolutions.

\subsection{Velocity structure}
\label{VS}
\begin{figure}
 \includegraphics[width=6.0cm,angle=-90 ]{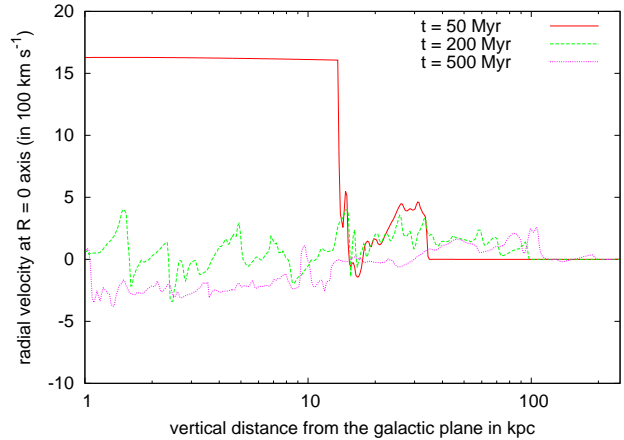}
 \caption {Velocity profiles along $R=0$ axis at 50, 200, 500 Myr for the case of $\mathcal{L}=10^{42}$ erg s$^{-1}$.}
 \label{velocity_snaps}
\end{figure}

\begin{figure}
 \includegraphics[width=6cm,angle=-90 ]{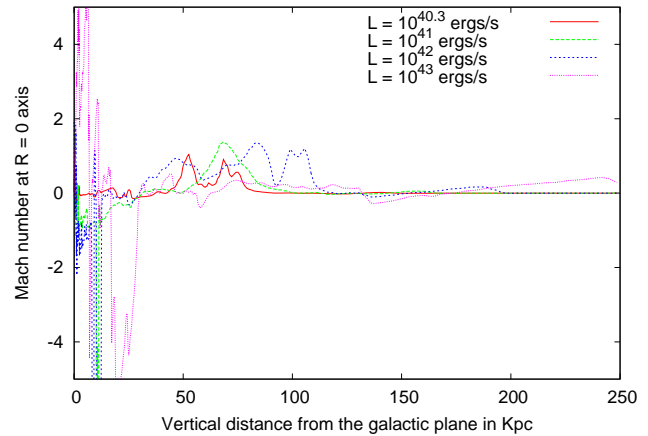}
 \caption {The Mach number profiles along the $R = 0$ axis at 500 Myr for different luminosities.
 }
 \label{vel_mach}
\end{figure}

Figure \ref{velocity_snaps} shows the velocity profiles at three different epochs (50,  200,  500 Myr) along the vertical direction
($R = 0$) for the fiducial run ($\mathcal{L}=10^{42}$ erg s$^{-1}$). The red solid curve for the profile at 50 Myr (when the injection is still on)
shows the structure of a standard luminosity-driven wind, with an inner region of free wind travelling at high speed, 
which is surrounded by the shocked wind, and then by the shocked
ISM, which drives an outer shock through the ambient medium. After this period (50 Myr), when the injection stops, the interaction zone produces
clumps which sometimes fall back and create regions with negative velocity. However,
the outer shock continues to propagate through the ambient
gas and reaches a distance of $200$ kpc in $500$ Myr in this case. The speed of the hot gas in the interior region depends crucially on the assumption
of the mass loading factor at injection, and in the case of $\dot{M}=0.1$ SFR, it reaches $1600$ km s$^{-1}$.
This is consistent with the analytical velocity of luminosity driven winds  ($\sqrt{2\mathcal{L}/\dot{M}}$; 
e.g., \citealt{sharma13a}, \citealt{CC85}).
We also note that, although the initial speed of the outflowing gas is $\sim1000$ km s$^{-1}$ or above, 
 it is
 not sustained for long and at later times when injection is turned off, the velocity becomes so small that it can be considered as a sound wave moving 
 through the hot medium. 
 This can be seen in Figure \ref{vel_mach} which plots the Mach number of the gas as a function of distance at 500 Myr for different luminosities.
 The Mach number of the outflowing gas at large distances decreases to $\lesssim 1$, for the mechanical luminosities considered here. Therefore the
 outflows eject the gas out of the virial radius with speeds comparable to the sound speed of the halo gas. This has important implications for the
 enrichment models of the IGM. Next, we focus on the wind structure at observable scales ($\sim 10$ kpc) based on our small-scale/short-duration simulations.

While ploughing through the ISM, the wind fluid  entrains the warm disk gas with it. For low
luminosities ($\lesssim 10^{41}$ erg s$^{-1}$)  this entrained gas mixes with the wind and forms filaments and
cloud-like structures embedded within the 10 kpc free wind. For higher SFR, the disk  gas is mainly located near the contact discontinuity of wind cone.  
While being carried away by the high velocity wind, a fraction of the cold clumps gets evaporated and the rest propagates outwards due 
to the ram pressure of the free wind. Therefore, the dynamics of the clouds and filaments is momentum conserving, for 
which the velocity increases with the distance \citep{murray05}. 
As the density of the hot gas decreases with distance, the ram pressure decreases, leading to an asymptotic
speed of the clouds. However, this result pertains to a steady state situation, which is not the case here. 
 The result obtained here is suitable for comparing the cloud kinematics at a particular time as obtained in observations.

\begin{figure}
 \includegraphics[trim=0.0cm 3.5cm 0cm 0cm, clip=true, totalheight=0.45\textheight,angle=-90]{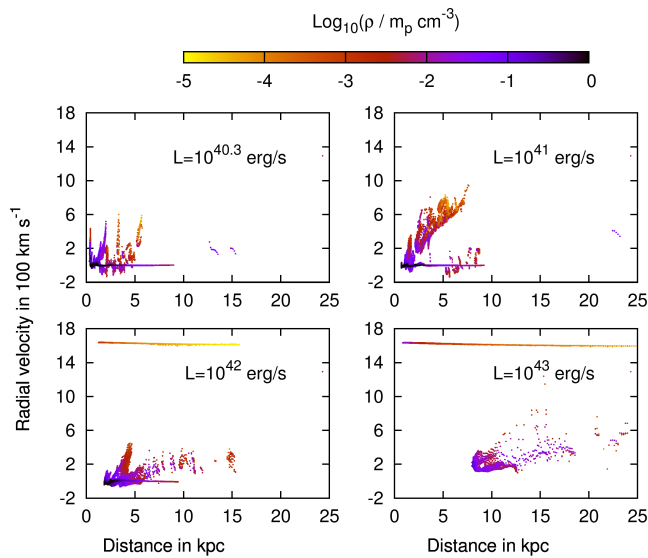}
 \caption {Scatter plot of the velocity of the warm gas ($T < 3\times 10^5\,$K) and radial distance at $50$ Myr. Top panel 
 		   is for $\mathcal{L} = 10^{40.3}$, $10^{41}$ erg s$^{-1}$ and bottom panel is for 
 		   $\mathcal{L} = 10^{42}$ and $10^{43}$ erg s$^{-1}$ respectively. 
	}
 \label{RV}
\end{figure}

 Figure \ref{RV} shows the position and velocity of warm/cold gas ($T< 3 \times 10^5$ K)  for four different
 luminosities at $50$ Myr. The figure shows that the velocity of the cold and warm gas ranges from  
 $\sim -150$ km s$^{-1}$ to
 $\sim 800$ km s$^{-1}$. The points with constant velocity  at $\simeq 1600$ km s$^{-1}$ represent the adiabatically
 cooled free wind in case of $\mathcal{L} = 10^{42}$ and $10^{43}$ erg s$^{-1}$, while the points with nearly 
 zero velocity represent the stationary disk  gas.

 For $\mathcal{L} = 10^{41, 42}$ erg s$^{-1}$, we also notice
  two sequences of velocity points, one which is a dominant sequence (referred to as the main sequence here),
  which extends from zero velocity to a velocity of $\sim 800$ km s$^{-1}$, 
  and, a secondary sequence which is almost parallel to the main sequence but extends from $-150$ km s$^{-1}$ to $+200$ km s$^{-1}$.
  Both sequences are almost linearly dependent on the radius. 
  This can be understood as the effect of ram pressure of the outgoing free/shocked wind, as mentioned previously.
  The radial dependence of the velocity of the warm gas in our simulation can be compared with the results obtained by \cite{shopbell98} 
  in case of H$_{\alpha}$ filaments in M82, who also observed a roughly linear relation between
  velocity and height above the disk (their Figure 10).

\begin{figure}
 \includegraphics[trim=4.0cm 3.5cm 0cm 0cm, clip=true, totalheight=0.4\textheight,angle=-90]{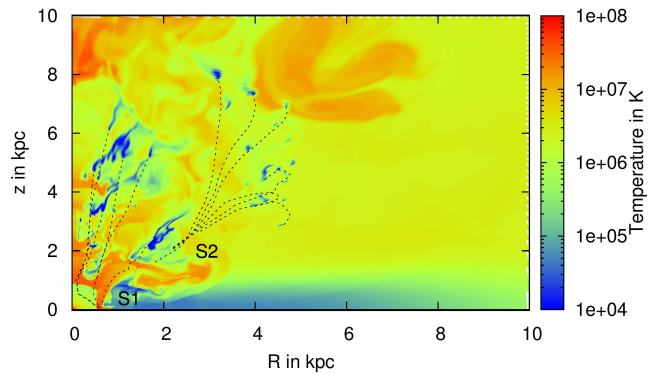}
 \caption {Temperature  map for $\mathcal{L}=10^{41}$ erg s$^{-1}$ at 50 Myr for, on which
 we superpose the tracks for cold clouds. The main sequence (S1) clouds (entrained by high
 velocity free wind) are tracked back for 17 Myr, and the secondary sequence (S2) clouds (entrained
 by the low velocity shocked wind) are tracked back for 40 Myr.}
 \label{tracks}
\end{figure}
 \begin{figure}
 \includegraphics[trim=2.0cm 2.0cm 0cm 0cm, clip=true, totalheight=0.4\textheight,angle=-90]{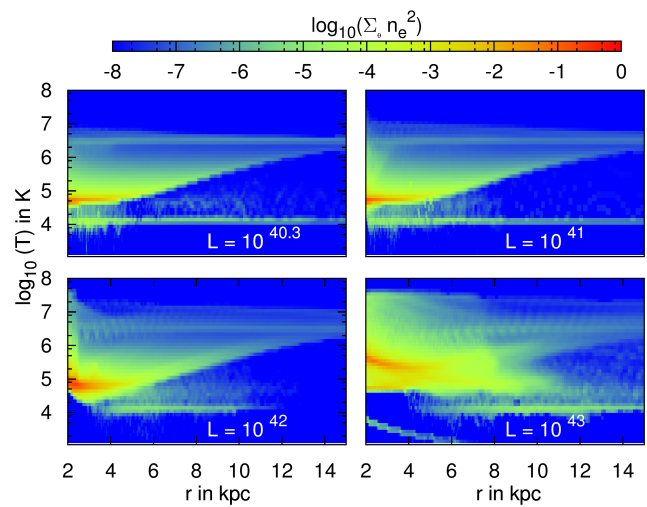}
 \caption {Scatter plot  of temperature and radial distance of gas particles, colour coded by the time average of
 density squared (averaged over $50$ Myr), shown for four different mechanical luminosities. 
           }
 \label{EM}
\end{figure}
 The origin of these two sequences are quite similar. The main sequence is entrained by the high velocity free wind, thereby giving
  it a relatively higher velocity. On the other hand, the secondary sequence arises because of the entrainment of the clouds by
  the lower velocity shocked wind. As shown in the evolution of wind for 
  $\mathcal{L}=10^{41}$ erg s$^{-1}$ in Figure \ref{contour41}, the main sequence corresponds to
  the clouds formed after the disk material advances inwards and is blown away into a filamentary
  structure (see snapshots at 40 and 50 Myr in Figure \ref{contour41}). The
  second sequence of clouds corresponds to the ones formed in the interaction zone between
  the hot halo gas and shocked ISM due to various instabilities  like thermal, Rayleigh-Taylor instabilities. (see Fig. \ref{contour0}).
   
   Figure \ref{RV} also shows the cloud gas density in colours. The clouds at large distances are in general more tenuous
  than those at inner region, which can be understood from adiabatic expansion of clouds moving in an ambient medium (free wind) whose
  pressure decreases with distance.

  The extension of the secondary sequence from $\sim -150$ km s$^{-1}$ to $+200$ km s$^{-1}$ means that few clouds are also 
  falling back to the centre. The fraction of mass that falls back to the center, however, is small to trigger any
  noticeable star formation, as will be discussed in the next section. Figure \ref{tracks} shows the time tracks of these two sequences of clouds
  seen at 50 Myr for $\mathcal{L}=10^{41}$ erg s$^{-1}$. The main sequence (labelled S1) are tracked back for 17 Myr, and represents relatively younger disk  material,
  whereas the secondary sequence clouds (S2) are tracked for 40 Myr, which are basically older population clouds.
  These two families of tracks clearly shows the source of the clouds and supports our previous discussion about their origin
  in the free and shocked wind.
  Other than these two sequences, we also notice some island points  (in Figure \ref{RV}) at a galacto-centric radius of $\sim 15\hbox{--}25$ kpc
  having velocity close to $\sim400$ km s$^{-1}$, which may represent rare high velocity-high latitude clouds as observed by \cite{sembach02}.
  
 Figure \ref{EM} shows the  temperature and position of gas parcels, colour coded by the time average
 of the square of particle
 density (averaged over $50$ Myr), for four different mechanical luminosities. Only the parcels of gas within $15$ kpc are
 represented here. The two horizontal streaks at $10^{6.5}$ K
 and $10^4$ K corresponds to the hot halo and
 warm clouds, respectively, whereas the rising envelope of increasing temperature with radial distance corresponds to the mixture of the disk and halo
 gas in the plane of the disk. The $1/r^2$ fall of temperature in case of $\mathcal{L} = 10^{43}$erg s$^{-1}$ is easily
 understood as the adiabatically expanding gas.
 The regions marked in red and orange  (back and deep gray in gray-scale) correspond to gas with high emissivity, and therefore are important from the 
 consideration of observability. The figure suggests that for very low luminosity
 outflows, most of the emission would arise from gas at $\sim 10^5$ K gas within $\sim 5$ kpc. X-ray emitting gas 
 becomes important for $\mathcal{L}\ge 10^{42}$ erg s$^{-1}$, corresponding to SFR of $\sim 10$ M$_\odot$ yr$^{-1}$. 
 These results are consistent with observations of X-rays from outflows \citep{strickland07}, 
 including the X-ray emission from the outflow  in Milky Way \citep{snowden95, breitschwerdt94}.
 
 \subsection{Mass inventory}
 In addition to the mass loading factor, the velocity and the temperature distribution, we have also 
 estimated the total outflowing mass. This is an important parameter in the context of the evolution of 
 the galactic disk and halo, as well as the enrichment of the IGM. The total mass injected into the halo 
 is assumed to be proportional to the SFR or $\mathcal{L}$  (Eq. \ref{mdotL2}), and it is a small fraction of 
 the gas mass in the halo, even for the largest SFR considered here. However, the total mass of the outflowing 
 gas ranges between $0.2\hbox{--}10 \%$ of the total gas content ($10^{11}$ M$_\odot$), increasing roughly linearly
 with SFR, between $1.5\hbox{--}150$ M$_\odot$ yr$^{-1}$. Therefore outflows corresponding to large SFR can change the halo gas 
 density by $\le 10\%$. We have also found that the average disk  mass does not change appreciably by either ejection or 
 fall back of gas ($\le 1\%$ for the most vigorous outflows) and in all cases, the change in the disk mass
 is much smaller than the injected mass.
 Previous works have discussed the role of the halo gas in massive galaxies either suppressing or triggering star
  formation in the disk by absorbing the outflowing gas or pushing it back on to the disk, respectively.The massive halo
  is expected to play an important role in quenching star formation in massive galaxies by cutting off the supply of fresh cooling gas.
  The halo also suppresses appreciable fall back
  of outflowing gas and quenches star formation, although the detailed mechanism of such quenching remains uncertain \citep{gabor14, oppenheimer10}.
 The above mass estimation implies that the injected material mostly gets
 deposited in the halo. Hence these outflows do not trigger further star formation by recycling mass to the disk (e.g., as in a galactic fountain).
 In other words, outflows in the presence of an extended hot halo gas can quench star formation in the galaxy.

\section{Discussion}
 We discuss a few implications of our results presented in the preceding sections.
\subsection{Definition of mass loading factor}
 We have defined the mass loading factor here as the ratio between the total outflowing mass and the total mass of stars formed.
This is in contrast with the usual definition, as the ratio between the mass outflow rate to the {\it current} SFR. Our definition is 
motivated by the fact that by the time the outflowing mass reaches the outer halo, its dependence on the {\it current} SFR loses its significance, 
since the duration of the SFR ($t_{\rm inj}$) is usually much smaller than $\sim 1$ Gyr, the time taken by the outflow to reach the virial radius. 
However, these two definitions are related to each other, and here we briefly discuss their inter-relation.

 These two definitions coincide if in the case of (a) outflows at small length scales and (b) when the starburst activity last for a  long time.
In the first case, the observed {\it current} SFR is related to the cause of the outflow. In the case, if the starburst activity lasts long time ($\gtrsim 300$ Myr) or there are
repeated bursts
at the centre, the outflow properties (viz. velocity, metallicity etc) at $\sim 100\hbox{--}200$ kpc also can be connected to the ongoing star formation process,
as observed by \cite{tumlinson2011}. Suppose one had defined the outer mass loading factor  as the ratio  between averaged mass outflow rate and 
the SFR as
\begin{equation}
 \label{eq:eta_defavg}
 \langle \eta_v \rangle = \frac{\langle \dot{M}_{\rm out} \rangle}{\rm SFR} = \frac{1}{\rm SFR \times t} \int_0^{t} \dot{M}_{\rm out} (r_v,t')\,dt'\,,
\end{equation}
where, $t$ is the averaging time, which can be taken roughly equal to the time taken by the shell to reach that particular radius. 
This definition of $\langle \eta_v \rangle$ can be connected to our earlier definition in Eq. \ref{eq:eta_def} as
\begin{equation}
\label{eq:two_etas}
 \eta_{v} = \frac{M_{\rm out}}{M_{\ast}^{+}} = \frac{\langle \dot{M}_{\rm out} \rangle}{\rm SFR} \times \frac{t}{t_{\rm inj}} 
 = \langle \eta_{v} \rangle \times \frac{t}{t_{\rm inj}} \,.
\end{equation}
Note that, this relation holds only when $t > t_{\rm inj}$. For $t<t_{\rm inj}$, the total mass of new stars formed is $M_{\ast}^{+} = \hbox{SFR} \times t$, therefore, $\eta_v = \langle \eta_v \rangle$.
The ratio $t/t_{\rm inj}$ can be estimated from our simulations as follows. The shell arrival  time in our simulation can be written as
$r_{sh} \approx 1.3\,\hbox{kpc}\,\mathcal{L}_{42}^{1/5}\,t_{\rm Myr}^{3/4}$ which in turn gives
$t/t_{\rm inj} \approx 17\,r_{\rm 200kpc}^{4/3}\mathcal{L}_{42}^{-4/15}t_{\rm inj, 50 Myr}^{-1}$\,. Therefore, Eq. \ref{eq:two_etas} gives us
\begin{equation}
 \eta_{v} \approx 17\,\langle \eta_{v} \rangle\,\,r_{\rm 200kpc}^{4/3}\mathcal{L}_{42}^{-4/15}t_{\rm inj, 50 Myr}^{-1}\,.
\end{equation}
From this equation we can clearly see that when the star formation lasts long i.e $t \approx t_{\rm inj}$, 
$\eta_{v} \approx \,\langle \eta_{v} \rangle$. In other words, the two definitions (one w.r.t. the average outflow rate, and another, 
presented here, w.r.t. the total outflowing mass) are equivalent in the case of long duration starbursts.

\subsection{Dust in clouds}
Clouds formed in the galactic outflows are not only important for containing
ions that make them observable, but they can also contain dust particles. Our result  shows that roughly half the outflowing mass
(inside 10 kpc) resides in gas of temperature $\sim 10^{6}$ K and the other half in warm clouds
of temperature $\sim 10^5$ K, has important implications for the types of dust particles that are likely to be embedded in outflows. 
The thermal sputtering rate of dust grains at $10^5$ K is small, and 
 the time scale required to destroy even the smallest dust grains ($\sim 0.003\, \mu$m) in warm clouds is $\sim 15$ Gyr
considering a density of $\sim 10^{-3}$ m$_p$ cm$^{-3}$ in these clouds, as inferred from the density distributions in Figures \ref{contour0}
and \ref{contour41}. These clouds can therefore preserve even the smallest grains, as long as the clouds can survive. The hotter regions
in which half the mass of the outflowing gas resides, has a larger sputtering rate. At $\sim 10^6$ K, the smallest grains that can survive
after 50 Myr is roughly $0.003$ $\mu$m for graphites and $0.03$ $\mu$m for silicates. These clouds therefore contain `grey' dust. 
In other words, half the dust mass carried by outflows are likely to be rendered 'grey' during the transport from the disk to the outer halo.

\subsection{Absorption study of clouds}
Our Galaxy has a SFR of $\approx 3$ M$_\odot$ yr$^{-1}$, which corresponds to a mechanical
luminosity of $\sim 10^{41.3}$ erg s$^{-1}$. Therefore the results of simulations with $\mathcal{L}=10^{41}$ erg s$^{-1}$ are
appropriate for comparison with our Galaxy. The numerous clouds that are formed during different stages would correspond to clouds
observed in various wavelengths in the halo of Milky Way. Cold clouds with $T \le 10^4$ K would correspond to HI clouds or MgII 
absorption clouds. The cold clouds seen in the temperature  distributions in Figure \ref{tracks} portray a visual impression of a likely 
scenario of clouds responsible for MgII absorption, although we emphasise that
we do not aim to reproduce the Milky Way observations in our paper. Shooting lines of sight from the centre in the range of 
$\theta=0\hbox{--}70^\circ$ (avoiding lines of sight within $20^\circ$ of the disk), we estimate a covering fraction of $\sim 60\%$ for
MgII clouds. This is consistent with the estimate of \cite{lehner12} for fraction of high velocity clouds with MgII, although the
correspondence should be interpreted with caution.

\subsection{Redshift dependence}
The specific star formation rate (sSFR, defined as the SFR per unit stellar mass) of Milky Way type galaxies increases at high redshift.
\cite{weinmann12} found the sSFR of galaxies with stellar mass $\sim 10^{10}$ M$_\odot$ to increase by a factor of $\sim 20$ at $z \sim 2$.
The corresponding star formation time scale (1/sSFR) decreases from the current value of $\sim 10$ Gyr to $\sim 0.5$ Gyr.
Therefore, the appropriate mechanical luminosity for counterparts of Milky Way at high redshift would be $\mathcal{L}\sim 10^{42.6}$ erg s$^{-1}$.
 As Fig. \ref{eta_L1} shows, for mechanical  luminosities of this order, the mass loading factor at the virial radius is close to unity.

\subsection{IGM enrichment}
The result that outflows leave the virial radius with a speed comparable to the sound speed of the halo gas may affect the enrichment
history of the IGM. The sound speed of the halo gas at virial temperature is roughly half the escape speed at the virial radius, over a
large range of masses and redshift. It is generally believed that the speed of the outflows is much larger than the escape speed. If the 
outflow speed is decreased as found here, then the radius of influence of the outflows in the IGM will be smaller than previously thought.
However, we should note that this result holds only for large galaxies with hot halo gas, whereas most of the contribution to the enrichment
of the IGM comes from low mass galaxies (e.g., \citealt{nath97, madau01, ferrara00b, oppenheimer06}), which may not harbour a hot gas in the halo.

\section{Summary}
\label{sec:summary}
In this paper, we have presented an extensive numerical study of SN driven galactic outflows for a MW type galaxy. Our modelled galaxy 
contains a gaseous disk of $T = 10^4$ K, and an extended hot ($T = 3 \times 10^6$ K) halo gas around it. The SN feedback was implemented in the 
form of mechanical energy within a compact region ($< 60$ pc) at the center of the galaxy. We have studied the effect of such energy inputs in small
scales ($\sim 30$ kpc) and in large scales ($\sim 200$ kpc). The small scale studies reveal the presence of multiphase structure of the outflowing
material and a temperature dependent outflow rate and thus help us to connect the mass loading factor at virial radius to the observable mass loading
factor.

We summarise our work as follows. 

(i) \textit{Mass loading factor}: The presence of hot halo gas in galaxies increases the mass loading factor compared to the no-halo case.
In the inner region (within $\sim 10$ kpc), the mass loading factor can
increase up to a factor of $\approx 5$  compared to $\eta_{\rm inj}$ when the mass fraction of the hot halo gas is $\sim 0.1$. In comparison, the 
mass loading factor near the virial radius ($\eta_v$), can  increase up to $10\hbox{--}40$ when compared with $\eta_{\rm inj}$. 
For low value of baryon fraction ($f_b$), the mass loading factor can be as low as $0.1$  (i.e. equal to the $\eta_{\rm inj}$).
The effect of the  halo gas in determining outflow rate is more pronounced in case of low star formation rates compared to the higher ones.

 Though we have mainly considered a single starburst of duration $50$ Myr, we have also shown that multiple bursts at the center that are well 
separated in time, have similar effects on gas at large radii. A comparative study with different star formation periods suggests that
$\eta_v$ depends only on the total mass of new stars formed. However, for higher mechanical luminosity 
($\mathcal{L} \gtrsim 10^{42}$ erg s$^{-1}$) cases, a star formation period longer than $\sim50$ Myr leads to significant radiative cooling and $\eta_v$
in this case also depend on the injection time scale.

We also found that the hot halo gas helps to quench star formation in the disk by inhibiting any appreciable recycling of 
mass into the disk.

(ii) \textit{Temperature distribution and observability}: The temperature distribution of the outflowing gas is approximately bimodal,
peaking at $10^{5}$ and $10^{6.5}$ K. This bimodality  implies that half of the outflowing gas is in form of warm clouds/gas
and other half is in the form of hot X-ray emitting gas. This result allows us to connect the  mass outflow rate for cold/warm gas to the 
outflow rate at the virial radius. We find that for a SFR of $1\,{\rm M}_{\odot}$ yr$^{-1}$, the  total (all of it in the hot phase)
mass loading factor at the virial radius is roughly 
 25 times the mass loading factor for cold/warm gas near the center  for a baryon fraction of 0.1 and injection time $\sim 50$ Myr.

(iii) \textit{Velocity}: The velocity of the free wind is found to be close to $\sim 1600$ km s$^{-1}$ when the energy injection is still active.
Velocity of the outflow decreases once the injection is switched off and for SFR $\lesssim 10$ M$_{\odot}$ yr$^{-1}$ after $\sim 400$ Myr, 
it becomes comparable to the sound speed
of the medium. Therefore, the velocity with which the outflow exits the galaxy is close to the sound speed of the hot medium ($c_s \sim 200$ km s$^{-1}$)
and the density of the outflow is also close to the halo medium ($\sim 10^{-4}$ m$_p$ cm$^{-3}$). We also notice that, even for a SFR
$\sim 150$ M$_{\odot}$~yr$^{-1}$, the whole galaxy is not 'blown away'. Due to the presence of the hot halo, a strong starburst acts like 
only a  perturbation at the center, after which, the galaxy relaxes and forms a disk -like structure again.

 (iv) \textit{Cloud velocity}: The velocity of the warm clouds in our small-scale simulations found to form two sequences in position-velocity diagram.
One, extends almost linearly from $0$ to $\sim 800$ km s$^{-1}$, another, extends from $\sim -200$ to $\sim +200$ km s$^{-1}$. These two sequences are signature of
entrainment of warm clouds by the high velocity free wind and the low velocity shocked wind respectively.

To conclude, our work focuses on the relation between the mass loading factor at various radii and connects them to the SFR without
coupling it to the total mass or  gas surface density of the galactic disk. 
Therefore, relations obtained in this paper can be used to extrapolate 
the observed outflow rate near the center to the outflow rate at the virial radius, for Milky Way type galaxies. A more general study of the dependence
of mass loading factor on the galactic mass will be addressed in a future study. 

\bigskip

\textbf{ACKNOWLEDGEMENT} 

We thank Romeel Dav\'e for his useful discussions.  We also thank anonymous referee for his comments.
KCS is partly supported by CSIR (grant no 09/1079(0002)/2012-EMR-I). 
 PS is partly supported by DST-India grant no. Sr/S2/HEP-048/2012. YS is supported by RFBR (grant no. 12-02-00917).

\footnotesize{

\appendix

\section{Rotation curve and gravitational acceleration}
\label{rotation_check}
For the mass model of the galaxy, we take $M_{\rm disk }$ = $5 \times 10^{10} {\rm M}_{\odot}$, $a = 4.0$ kpc, $b = 0.4$ kpc \citep{smith07}, $c = 12.0$ 
\citep{maccio07}
, $d = 6.0$ kpc (the flattening length) and $M_{\rm vir}$ = $10^{12} {\rm M}_{\odot}$, which gives us $r_{\rm vir} = 258$ kpc and 
$r_s = 21.5$ kpc. We note that \cite{smith07} used a rotation velocity  $\simeq 220$ km s$^{-1}$
with $c = 24$, and our assumption of $c=12$ decreases the rotation velocity by $15\%$, which we consider
to be negligible. The rotation curve can be found from the equation
\begin{align}
\frac{v_T^2}{R} &= \left[\frac{\partial \Phi}{\partial R} \right]_{z=0} \\
                &= \left[\frac{\partial }{\partial R}(\Phi_{\rm DM}+\Phi_{\rm MN}) \right]_{z=0}  \nonumber
\end{align}
which gives (in units of $10^{14}$ (cm s$^{-1}$)$^2 $)
\begin{align}
v_T^2 &= -\frac{12.3\,R^2}{\left(d^2+R^2\right) \left(1+\frac{\sqrt{d^2+R^2}}{r_s}\right)} + \frac{21.5\,R^2}{\left(R^2+\left(a+b\right)^2\right)^{3/2}}  \\
      &+ \frac{12.3\,R^2\,r_s \log\left[1+\frac{\sqrt{d^2+R^2}}{r_s}\right]}{\left(d^2+r^2\right)^{3/2}}  \nonumber
\end{align}
The rotation curve is shown in Figure \ref{rotation_curve}.

\begin{figure}
\includegraphics[width=8.4cm]{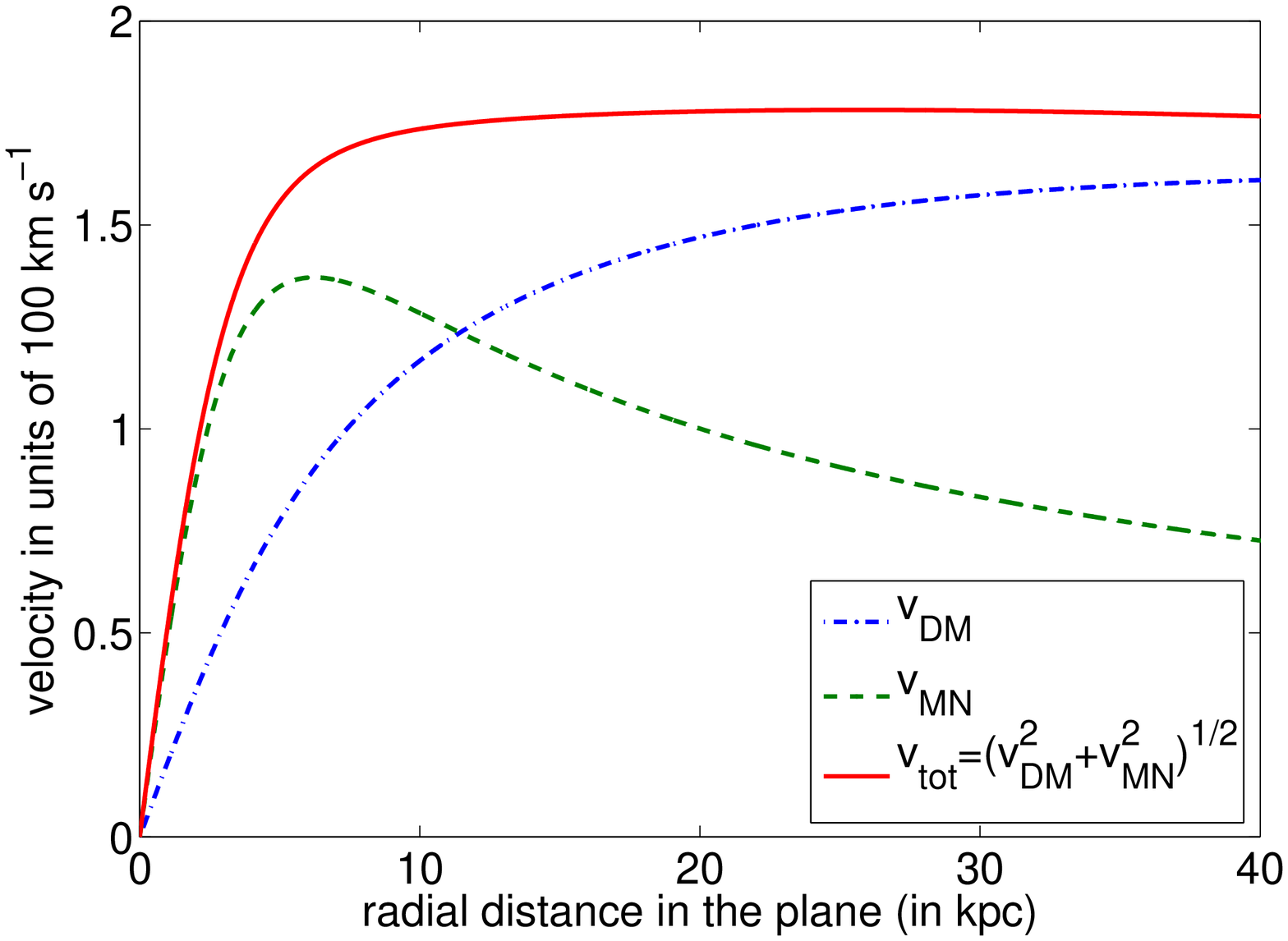}
\caption{
The rotation curve for the galaxy. The plot also shows the contributions from different potential components.
}
\label{rotation_curve}
\end{figure}

To visualise the gravitational acceleration ($g$) in the plane and perpendicular to the plane , we plot different components of $g$ in Figure \ref{gravity}.
These curves show that the acceleration is not constant in the central region of the Galaxy. This 
happens because when we go up in $z$-direction, the total gravitating mass, which can influence a test particle, increases with 
height. However, at a certain height contribution from the stellar disk  becomes maximum, and after that distance, the gravitating mass does
not increase much and the acceleration decreases purely due to distance effect. 

\begin{figure}
 \centerline{
 \includegraphics[width=4.6cm]{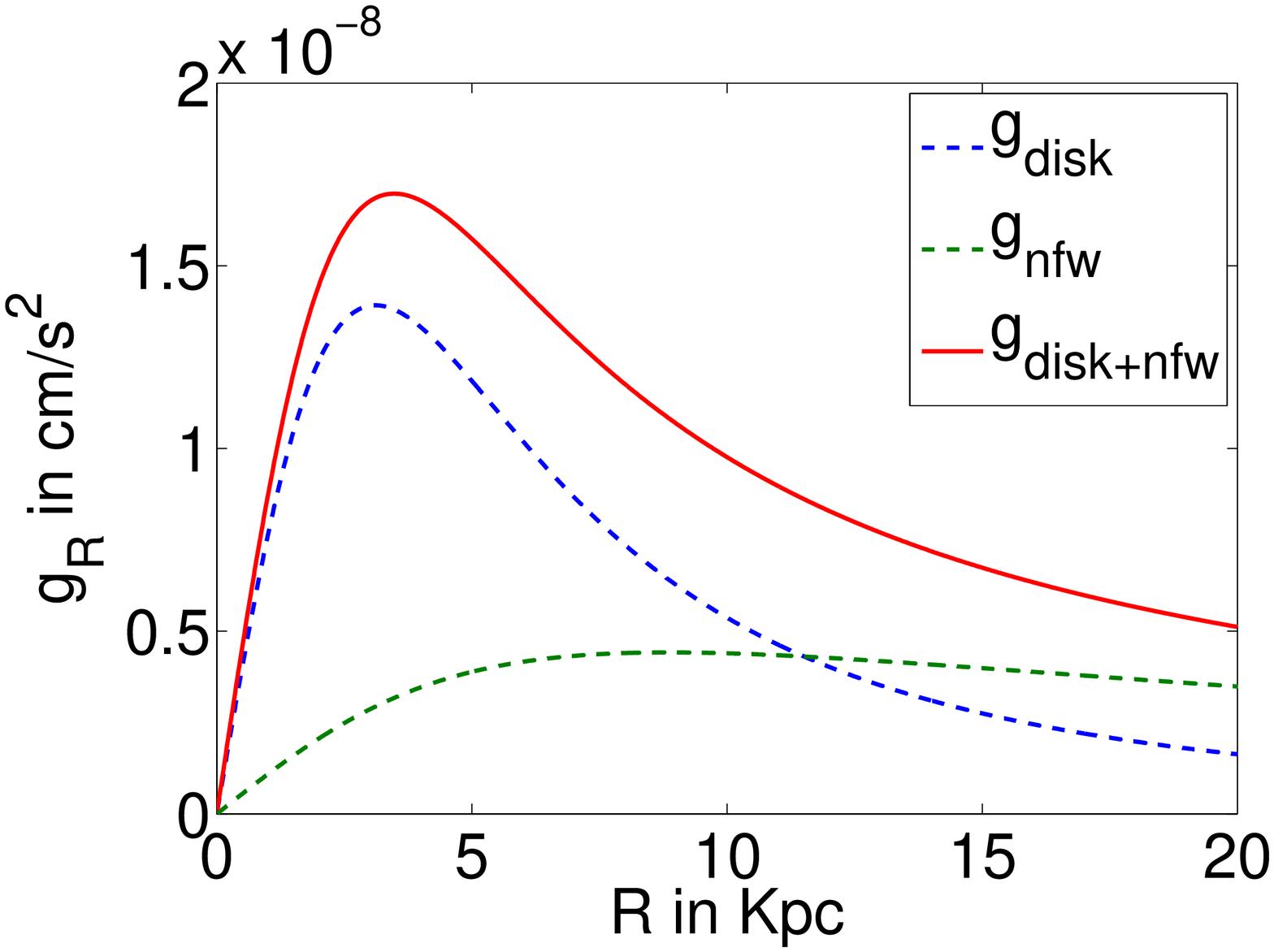}
 \includegraphics[width=4.5cm]{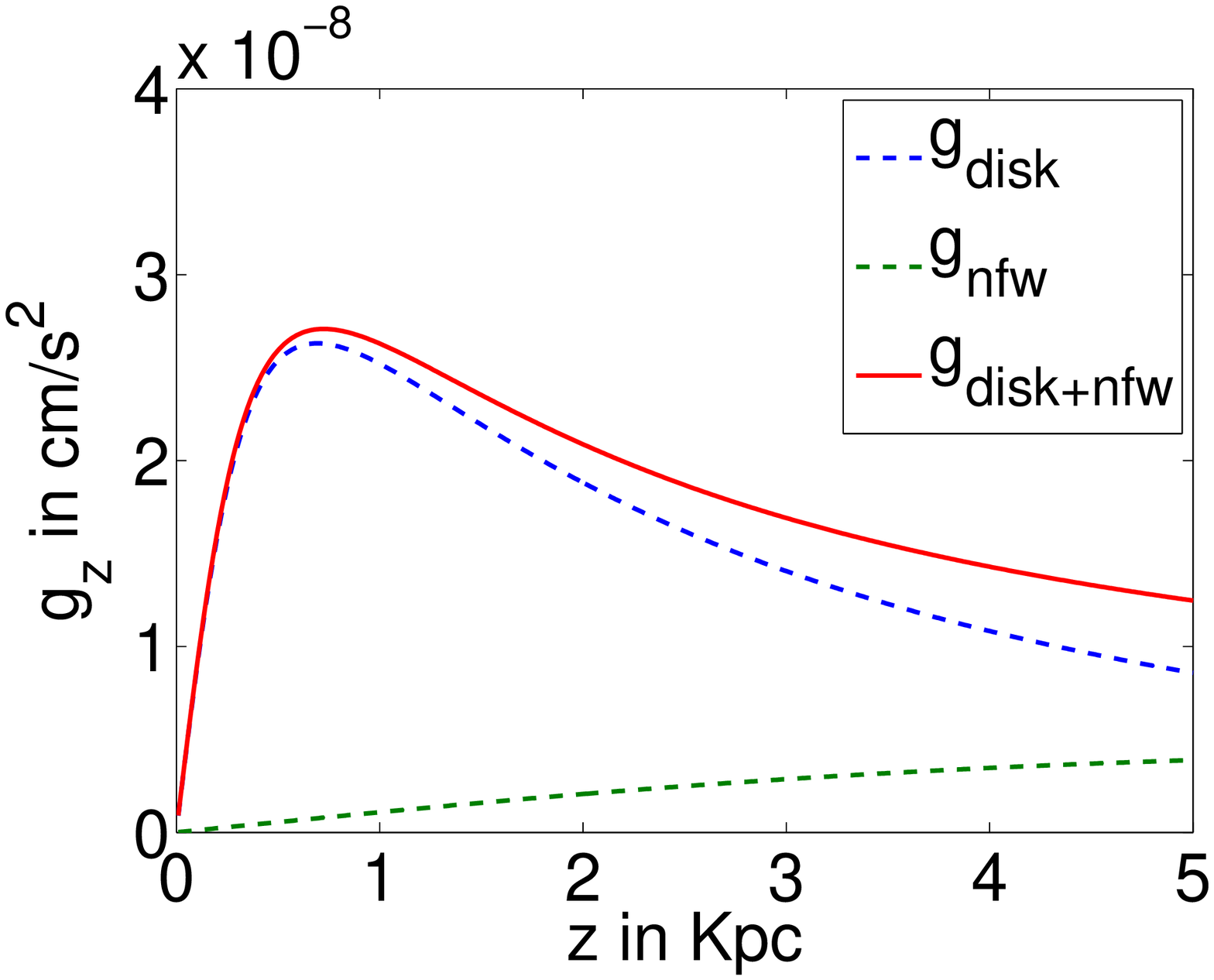}
 }
 \caption{The gravitational acceleration in the plane (left panel) and perpendicular to the plane (right panel).}
          \label{gravity}
\end{figure}

\section{ Modified NFW profile }
\label{DM}
The modified form of NFW profile considered by us is 
\begin{equation}
  \Phi_{\rm DM} = - \left( \frac{GM_{\rm vir}}{f(c)\,r_s} \right) \frac{\log(1+\sqrt{R^2+z^2+d^2}/r_s)}{\sqrt{R^2+z^2+d^2}/r_s} \,,
\end{equation}
which avoids a singularity at the center. The dark matter density distribution of such a potential 
can be found using the Poisson's equation,
\begin{align}
 \rho_{\rm DM} (r) &= \frac{1}{4\pi\,G} \nabla^2\Phi_{\rm DM}  \nonumber \\
          &= \frac{1}{4\pi G} \frac{1}{r^2} \frac{d}{d r} \left(r^2\frac{d}{d r} \Phi_{\rm DM} \right) 
\end{align}
which gives 
\begin{align}
K\,\rho_{\rm DM}(r) &= \frac{3 r^2}{\left(d^2+r^2\right)^2 \left(1+\frac{\sqrt{d^2+r^2}}{r_s}\right)}
                                  -\frac{3}{\left(d^2+r^2\right) \left(1+\frac{\sqrt{d^2+r^2}}{r_s}\right)} \nonumber \\ 
                               &+  \frac{r^2}{\left(d^2+r^2\right)^{3/2} \left(1+\frac{\sqrt{d^2+r^2}}{r_s}\right)^2 r_s}
                                  -\frac{3 r^2 r_s \log\left[1+\frac{\sqrt{d^2+r^2}}{r_s}\right]}{\left(d^2+r^2\right)^{5/2}}  \nonumber \\
                               &+  \frac{3r_s \log\left[1+\frac{\sqrt{d^2+r^2}}{r_s}\right]}{\left(d^2+r^2\right)^{3/2}} \, .
\end{align}
where, $K = \frac{4\pi\,f(c)\,r_s}{M_{\rm vir}}$. This equation reduces to the standard NFW DM density distribution for $d = 0$ .
The density distribution of the dark matter can be shown in Figure \ref{DM_density}. 
\begin{figure}
   \includegraphics[width=8.4cm,height=6.3cm]{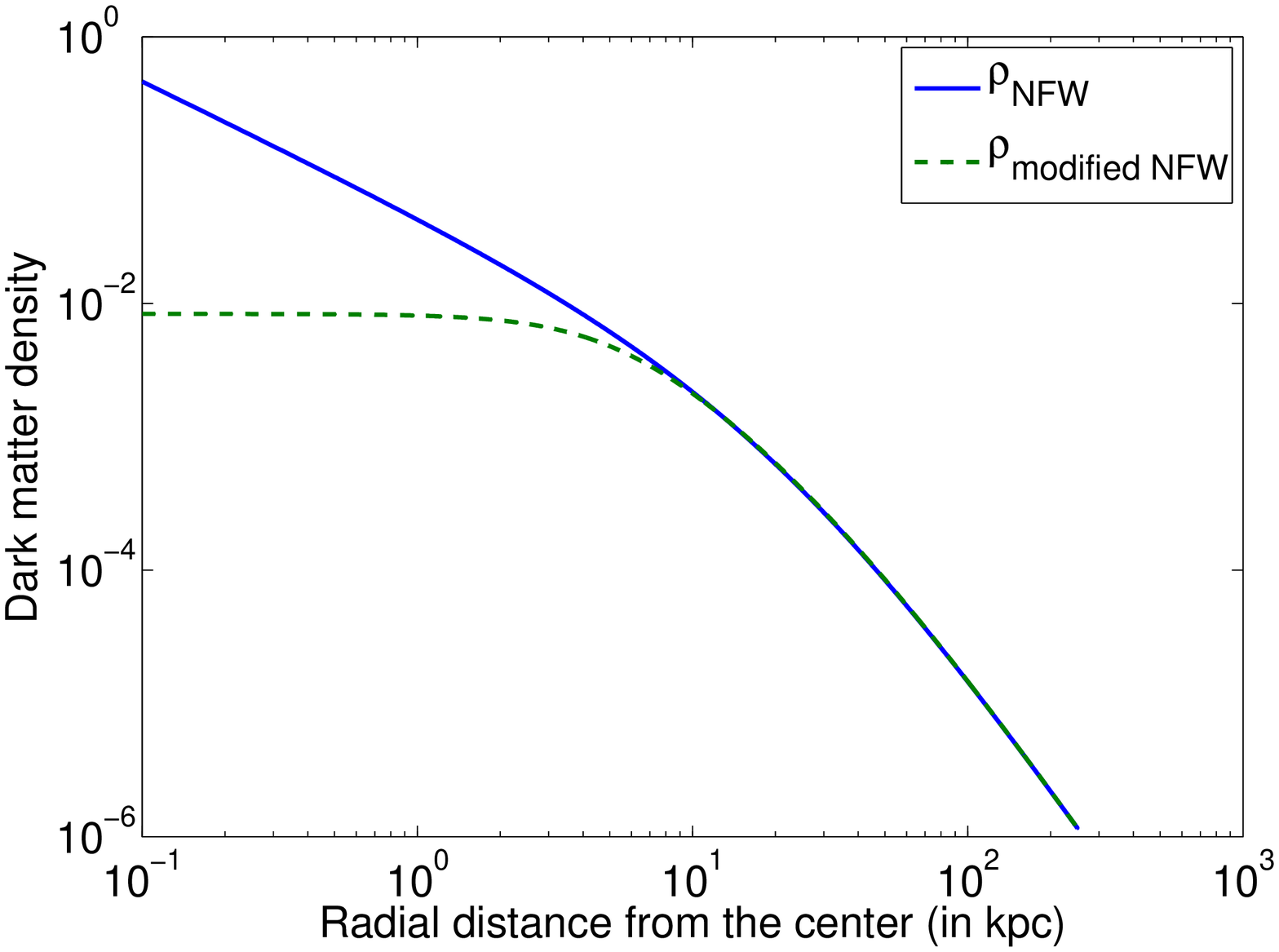}
   \caption{ The dark matter density distribution in the Galaxy in units of $\frac{M_{\rm vir}}{4\pi\,f(c)\, r_s}$. 
            The plot uses the parameters for the MW type galaxy mentioned in Table \ref{massparameters}.
            }
  \label{DM_density}
\end{figure}

\section{ Initial density setup}
\label{densitysetup}
In steady state, for a rotating fluid, force balance along $R$ and $z$ gives
\begin{eqnarray}
\label{c1}
 0 &=& -\frac{\partial \Phi}{\partial R} - \frac{1}{\rho} \frac{\partial p}{\partial R} + \frac{v_{\phi,g}^2}{R} \\
 \label{c2}
 0 &=& -\frac{\partial \Phi}{\partial z} - \frac{1}{\rho} \frac{\partial p}{\partial z} \, .
\end{eqnarray}
Here, we use $v_{\phi,g}(R)$ (gas angular velocity is a function of $R$ only) as the rotational velocity of the 
gas ($p\neq0$) compared to $v_{\phi,G}=\sqrt{R\frac{\partial \Phi}{\partial R}}$, 
which we use for the rotational velocity for a pressure-less test particle at $z=0$. Eq. \ref{c2} can be solved as 
\begin{eqnarray}
 \frac{kT}{\mu m_p \rho} \frac{\partial \rho}{\partial z}  &=& -\frac{\partial \Phi}{\partial z} \nonumber \\
 \log \rho &=& -\frac{\mu m_p}{kT} \Phi + f_1(R) \nonumber \\
 \label{c3}
 \rho(R,z) &=& \rho_{0R}(R) \exp\left( -\frac{\mu m_p}{kT} \Phi \right)  \,, 
\end{eqnarray}
where, $\rho_{0R}(R) = \exp[f_1(R)]$.
From Eq. \ref{c1}, we write,
\begin{eqnarray}
 \frac{kT}{\mu m_p \rho} \frac{\partial \rho}{\partial R}  &=& -\frac{\partial \Phi}{\partial R} + \frac{v_{\phi,g}^2}{R}   \nonumber \\
 \log \rho &=& -\frac{\mu m_p}{kT} \left( \Phi - \int \frac{v_{\phi,g}^2}{R} dR \right) + f_2(z)
\end{eqnarray}
  Eq. \ref{c3} then gives
  \begin{eqnarray}
 -\frac{\mu m_p}{kT} \Phi + \log \rho_{0R} &=& -\frac{\mu m_p}{kT} \left( \Phi - \int \frac{v_{\phi,g}^2}{R} dR \right)  + f_2(z) ,\nonumber \\
 \label{c4}
 \rho_{0R}(R) &=& \exp \left(  \frac{\mu m_p}{kT} \int \frac{v_{\phi,g}^2}{R} dR \right) \times {\rm constant}. \nonumber \\
\end{eqnarray}
 Since, $\rho_{0R}$ is only a function of $R$ by definition, $f_2(z)$ must be a constant.
Combining equations \ref{c3} and \ref{c4}, we get, 
\begin{equation}
\label{c5}
 \rho(R,z) = {\rm constant} \times \exp\left( -\frac{\mu m_p}{kT} \left[ \Phi -\int \frac{v_{\phi,g}^2}{R}\, dR \right]\right)\, .
\end{equation}
For simplicity, let us assume that the gas rotation velocity is a fraction of the particle rotation on the plane ($z = 0$), i.e. $v_{\phi,g} = f\,v_{\phi,G}$,
where, $f$ is a constant. 
Then, Eq. \ref{c5} can be written as 
\begin{equation}
\label{c6}
  \rho(R,z) = {\rm constant} \times \exp\left( -\frac{1}{c_s^2} \left[ \Phi(R,z) - f^2\Phi(R,0) \right]\right)\, .
\end{equation}
Here, $c_s = \sqrt{\frac{kT}{\mu m_p}}$ is the isothermal sound speed of the gas. For a non-rotating gas, the equation becomes 
\begin{equation}
\label{c7}
   \rho(R,z) = {\rm constant} \times \exp\left( -\frac{1}{c_s^2} \left[ \Phi(R,z) \right]\right)\, .
\end{equation}
The constant can be determined by  normalising the density.

Several isothermal components can be combined, as we do for a rotating WIM disc and a non-rotating hot halo.


\begin{thebibliography}{}

\bibitem[\protect\citeauthoryear{Arribas \etal}{2014}]{arribas14}
Arribas, S., Colina, L., Bellocchi, E., Maiolino, R., Villar-Mart\'in, M., 2014,  arXiv:1404.1082v1  


\bibitem[\protect\citeauthoryear{Babul \& Rees}{1992}]{babul92}
Babul, A., Rees, M. J., 1992, MNRAS, 255, 346
 
\bibitem[\protect\citeauthoryear{Bogdan \etal} {2013}]{bogdan13a}
Bogd\'an, \'A., Forman, W. R., Vogelsberger, M., Bourdin, H., Sijacki, D.,
Mazzotta, P., Kraft, R., Jones, C., Gilfanov, M., Churazov, E., David, L., 2013, ApJ, 772, 97 

\bibitem[\protect\citeauthoryear{Bogdan \etal}{2013}]{bogdan13b}
Bogd\'an, \'A., Forman, W. R., Kraft, R. P., Jones, C., 2013, ApJ, 772, 98
 
\bibitem[\protect\citeauthoryear{Bolatto \etal} {2013}]{bolatto2013}
Bolatto, A. D, Warren, S. S., Leroy, A. K., Walter, F., Veilleux, S., Ostriker, E., Ott, J.,
Zwaan, M. \etal 2013, Nature, 499, 450


 
\bibitem[\protect\citeauthoryear{Bordoloi \etal} {2014}]{bordoloi2014} 
Bordoloi, B., Tumlinson, J., Werk, J., \etal 2014, arxiv: 1406.0509 
 
\bibitem[\protect\citeauthoryear{Bouch\'e \etal} {2012}]{bouche2012}
Bouch\'e, N., Hohensee, W., Vargas, R., Kacprzak, G. G., Martin, C. L., Cooke, J., 
Churchill, C. W. 2012, MNRAS, 426, 801


\bibitem[\protect\citeauthoryear{Breitschwerdt \& Schmutzler} {1994}]{breitschwerdt94}
Breitschwerdt, D, Schmutzler, T. 1994, Nature, 371, 774

\bibitem[\protect\citeauthoryear{Chattopadhyay \etal}{2012}]{indranil12}
Chattopadhyay I., Sharma M., Nath B., Ryu D., 2012, MNRAS, 423, 2153

\bibitem[\protect\citeauthoryear{Chevalier \& Clegg} {1985}]{CC85}
Chevalier R. A., Clegg A. W., 1985, Nature, 317, 44



\bibitem[\protect\citeauthoryear{Cooper \etal} {2008}]{cooper08}
Cooper J. L., Bicknell G. V., Sutherland R. S., 2008, ApJ, 674, 157




\bibitem[\protect\citeauthoryear{Dalla Vecchia \etal} {2008}]{dalla08}
Dalla Vecchia C., Schaye J., 2008, MNRAS, 387, 1431

\bibitem[\protect\citeauthoryear{Dekel \& Silk} {1986}]{dekel86}
Dekel, A., Silk, J. 1986, ApJ, 303, 39



\bibitem[\protect\citeauthoryear{Fang \etal}{2013}]{fang13}
Fang, T., Bullock, J., Boylan-Kolchin, M., 2013, ApJ, 762, 20

\bibitem[\protect\citeauthoryear{Ferrara \etal} {2000a}]{ferrara00a}
Ferrara, A., Pettini, M., Shchekinov, Yu. A. 2000, MNRAS, 319, 539

\bibitem[\protect\citeauthoryear{Ferrara \etal} {2000b}]{ferrara00b}
Ferrara A., Tolstoy E., 2000, MNRAS, 313, 291

\bibitem[\protect\citeauthoryear{Fraternali \etal} {2006}] {fraternali06}
Fraternali F., Binney J. J., 2006, MNRAS, 449, 449


\bibitem[\protect\citeauthoryear{Gabor \& Dav\'e} {2014}]{gabor14}
Gabor, J. M., Dav\'e, R., 2014, arxiv:1405.1043v1

\bibitem[\protect\citeauthoryear{Gatto \etal}{2013}]{gatto13}
Gatto, A., Fraternali, F., Read, J. I., Marinacci, F., H. Lux, Walch, S., 2013, arxiv:1305.4176v2


\bibitem[\protect\citeauthoryear{Heald \etal}{2006}]{heald06}
Heald G. H., Rnad R. J., Benjamin R. A. and Bershady M. A., 2006, ApJ, 647, 1018

\bibitem[\protect\citeauthoryear{Heckman \etal} {2000}]{heckman00}
Heckman, T. M., Lehnert, M. D., Strickland, D. K., Armus, L. 2000, ApJS, 129, 493


\bibitem[\protect\citeauthoryear{Hopkins \etal} {2012}]{hopkins2012}
Hopkins P. F., Quataert E., Murray N., 2012, MNRAS, 421, 3522

\bibitem[\protect\citeauthoryear{Hopkins \etal} {2013} ]{hopkins2013}
Hopkins P. F., Keres D., Onorbe J., Faucher-Giguere C-A., Quataert E., Murray N., Bullock J. S., in press , arXiv: 1311.2073v1




\bibitem[\protect\citeauthoryear{Kalbera \etal}  {2009} ] {kalbera09}
Kalbera P. M. W., Kerp J., 2009, ARAA, 47, 27

\bibitem[\protect\citeauthoryear{Koyama \& Inutsuka} {2004}]{koyama04}
Koyama H., Inutsuka S., 2004, ApJ, 602, 25


\bibitem[\protect\citeauthoryear{Lagos \etal} {2013}]{lagos2013}
Lagos, C. del P. , Lacey, C. G., Baugh, C. M. 2013, 436, 1787


\bibitem[\protect\citeauthoryear{Larson}{1974}]{larson74} 
Larson, R. 1974, MNRAS, 169, 229

\bibitem[\protect\citeauthoryear{Lehner \etal} {2012}]{lehner12}
Lehner, N., Howk, J. C., Thom, C., Fox, A. J., Tumlinson, J., Tripp, T. M., Meiring, J. D. 2012,
MNRAS, 424, 2896

\bibitem[\protect\citeauthoryear{Leitherer \etal}{1999}]{leitherer99}
Leitherer, C., Schaerer, D., Goldader, J. D., Deldado, R. M. G., Robert, C., Kune, D. F., De Mello, D. F., Devost, D.
and Heckman, T. M., 1999, ApJSS, 123, 3

\bibitem[\protect\citeauthoryear {Liou }{1996}]{liou96}
Liou M. S., 1996, J. Comput. Phys., 129, 364



\bibitem[\protect\citeauthoryear{Mac Low \& Ferrara} {1999}]{maclow99}
Mac Low M-M., Ferrara A., 1999, ApJ, 513, 142

\bibitem[\protect\citeauthoryear{Maccio \etal}{2007}]{maccio07}
Maccio A. V., Dutton A. A., Bosch F. C. van den, Moore B., Potter D., Stadel J., 2007, MNRAS, 378, 55

\bibitem[\protect\citeauthoryear{Madau \etal}{2001}]{madau01}
Madau, P., Ferrara, A., Rees, J., M., 2001, ApJ, 555, 92

\bibitem[\protect\citeauthoryear{Martin} {1999}]{martin99}
Martin, C. 1999, ApJ, 513, 156


\bibitem[\protect\citeauthoryear{Martin \etal} {2012}]{martin2012}
Martin, C. L., Shapley, A. E., Coil, A. L., Kornei, K. A.,  Bundy, K., Weiner, B. J.,
Noeske, K. G., Schiminovich, D. 2012, ApJ, 760, 127

\bibitem[\protect\citeauthoryear{Mathes \etal} {2014}]{mathes2014}
Mathes, N. L., Churchill, C. W., Kacprzac, G. G., \etal 2014, ApJ, 792, 128 

\bibitem[\protect\citeauthoryear{Mcmillan \etal} { 2011}] {mcmillan11}
McMillan P. J., arXiv: 1102.4340v1

\bibitem[\protect\citeauthoryear{Melioli \etal}{2013}] {melioli13}
Melioli C., Dal Pino E. M. G., Geraissate F. G., arXive:1301.5005v1

\bibitem[\protect\citeauthoryear{Mignone \etal} { 2007}]{mignone07}
Mignone A., Bodo G., Massaglia S., Matsakos T., Tesileanu O., Zanni C., Ferrari A., 2007, ApJSS, 170, 228

\bibitem[\protect\citeauthoryear{Miyamoto \etal}  {1975}]{miyamoto75}
Miyamoto M., Nagai R., 1975, PASJ, 27, 533

\bibitem[\protect\citeauthoryear{Mo, Mao and White}{1998}]{momao98}
Mo H. J., Mao S., White S. D. M., 1998, MNRAS, 297, 319-336

\bibitem[\protect\citeauthoryear{Murray \etal} {2005}]{murray05}
Murray, N., Quataert, E., Thompson, T. A. 2005, ApJ, 618, 569


\bibitem[\protect\citeauthoryear{Nath \& Trentham} {1997}]{nath97}
Nath, B. B., Trentham, N. 1997, MNRAS, 291, 505


\bibitem[\protect\citeauthoryear{Nath \& Shchekinov} {2013}]{nath13}
Nath, B. B., Shchekinov, Y. 2013, ApJL, 777, 12

\bibitem[\protect\citeauthoryear{Newman \etal} {2012}]{newman2012}
Newman, S. F., Genzel, R., F\"orster-Schreiber, N. M., Shapiro Griffin K., Mancini M.,
Lilly, S. J., Renzini, A., Bouch\'e, N. \etal 2012, ApJ, 761, 43

\bibitem[\protect\citeauthoryear{Navarro \etal} {1996}]{nfw96}
Navarro J. F., Frenk C. S., White S. D. M., 1997, ApJ, 490, 493


\bibitem[\protect\citeauthoryear{Oppenheimer \etal}{ 2006}]{oppenheimer06}
Oppenheimer, B. D., Dave, R., 2006, MNRAS, 373, 1265

\bibitem[\protect\citeauthoryear{Oppenheimer \etal}{ 2010}]{oppenheimer10}
Oppenheimer, B. D., Dave, R., Keres, D., Fardal, M., Katz, N., Kollmeier, J. A., Weinberg, H., 2010, MNRAS, 406, 2325




\bibitem[\protect\citeauthoryear{Roy \etal}{ 2013}]{arpita13}
Roy A., Nath B. B., Sharma P., Shchekinov Y., 2013, 434, 3572



\bibitem[\protect\citeauthoryear{Rupke \& Veilleux} {2013}]{rupke2013}
Rupke, D. S., Veilleux, S., 2013, ApJ, 768, 75


\bibitem[\protect\citeauthoryear{Sembach \etal}{2002}]{sembach02}
Sembach, K. R., Wakker, B. P., Savage, B. D., Richter, P., Meade, M., 
Shull, J. M., Jenkins, E.B., Sonneborn, G., Moos, H. W., 2002, arxiv: 0207562v1

\bibitem[\protect\citeauthoryear{Sharma \etal}{ 2012}]{sharma12}
Sharma, P., McCourt, M., Parrish, I. J., Quataert, E., 2012, MNRAS, 427, 1219

\bibitem[\protect\citeauthoryear{Sharma \& Nath} {2013}]{sharma13a}
Sharma M., Nath B., 2013, ApJ, 763, 17

\bibitem[\protect\citeauthoryear{Sharma \etal}{2014}]{sharma13b}
Sharma, M., Nath B., Chattopadhyay I., Shchekinov Y., arXiv: 1306.4362v1

\bibitem[\protect\citeauthoryear{Sharma \etal}{ 2014}]{psharma14}
Sharma, P., Roy, A., Nath, B. B., Shchekinov Y., 2014, arXiv 1402.6695

\bibitem[\protect\citeauthoryear{Shopbell \& Bland-Hawthorn}{1998}]{shopbell98}
Shopbell, P. L., Bald-Hawthorn, J., 1998, ApJ, 493, 129

\bibitem[\protect\citeauthoryear{Silich \& Tenorio-Tagle}{2001}]{silich01}
Silich, S., Tenorio-Tagle, G., 2001, ApJ, 552, 91

\bibitem[\protect\citeauthoryear{Snowden \etal} {1995}]{snowden95}
Snowden, S. L. \etal 1995, ApJ, 454, 643

\bibitem[\protect\citeauthoryear{Springel \& Hernquist} {2003}]{sh03}
Springel, V., Hernquist, L., 2003, MNRAS, 339, 289

\bibitem[\protect\citeauthoryear{Strickland \& Stevens} {2000}]{strickland00}
Strickland D. K. and Stevens I. R., 2000, MNRAS, 314, 511

\bibitem[\protect\citeauthoryear{Stickland \& Heckman}{2007}]{strickland07}
Strickland, D. K. and Heckman, T. M., 2007, ApJ, 658,258

\bibitem[\protect\citeauthoryear{Smith \etal} { 2007}]{smith07}
Smith M. C., Ruchti G. R., Helmi A., Wyse R. F. G. \etal, MNRAS, 2007, 379, 755

\bibitem[\protect\citeauthoryear{Suchkov \etal}{1994}]{suchkov94}
Suchkov, A. A., Balsara, D. S., Heckman, T. M., Leitherer, C., 1994, ApJ, 430, 511 

\bibitem[\protect\citeauthoryear{Suchkov \etal}{1996}]{suchkov96}
Suchkov, A. A., Berman, V. G., Heckman, T. M., Balsara, D. S., 1996, ApJ, 463, 528

\bibitem[\protect\citeauthoryear{Sutherland \& Dopita}{ 1993}]{sutherland93}
Sutherland R. S., Dopita M. A., 1993, ApJSS, 88, 253

\bibitem[\protect\citeauthoryear{Swaters \etal}{1997}]{swaters97}
Swaters R. A., Sancisi R. and Van der Hulst J. M., 1997, ApJ, 491, 140


\bibitem[\protect\citeauthoryear{Tegmark \& Silk} {1993}]{tegmark93}
Tegmark, M., Silk, J. 1993, ApJ, 417, 54

\bibitem[\protect\citeauthoryear{Tumlinson \etal} {2011}]{tumlinson2011}
Tumlinson, J, Thom, C., Werk, J. \etal 2011, Science, 334, 948



\bibitem[\protect\citeauthoryear{Vasiliev \etal}{2014}]{vasiliev14}
Vasiliev E. O., Nath B. B., Bondarev R., Shchekinov Y., arXiv: 1401.5070v1


\bibitem[\protect\citeauthoryear{Weaver \etal} { 1977}]{weaver77}
Weaver R., McCray R., Castor J., Shapiro P., Moore R., 1977, ApJ, 218, 377

\bibitem[\protect\citeauthoryear{Weinmann \etal} {2012}]{weinmann12}
Weinmann, S. M. \etal 2012, MNRAS, 426, 2797

 



 
\end{thebibliography}
\end{document}